\documentclass[iopscience,apj,twocolappendix]{emulateapj}

\usepackage{placeins}
\usepackage{natbib}
\usepackage{amsmath}
\usepackage{color}

\begin{document}

\newcommand\be{\begin{equation}}
\newcommand\en{\end{equation}}
\newcommand\pdv{$P{\rm d}V~$}

\shorttitle{Spiral Wave Instability in Disks} 
\shortauthors{Bae et al.}

\title{SELF-DESTRUCTING SPIRAL WAVES: GLOBAL SIMULATIONS OF A SPIRAL WAVE INSTABILITY IN ACCRETION DISKS}

\author{Jaehan Bae\altaffilmark{1},
Richard P. Nelson\altaffilmark{2},
Lee Hartmann\altaffilmark{1},
Samuel Richard\altaffilmark{2}}

\altaffiltext{1}{Department of Astronomy, University of Michigan, 1085 S. University Ave.,
Ann Arbor, MI 48109, USA} 
\altaffiltext{2}{Astronomy Unit, Queen Mary University of London, Mile End Road, London E1 4NS, UK}

\email{jaehbae@umich.edu, r.p.nelson@qmul.ac.uk, lhartm@umich.edu, samuel.richard@qmul.ac.uk}

\begin{abstract}
We present results from a suite of three-dimensional global hydrodynamic simulations which show that spiral density waves propagating in circumstellar disks are unstable to the growth of a parametric instability that leads to break-down of the flow into turbulence. 
This spiral wave instability (SWI) arises from a resonant interaction between pairs of inertial waves, or inertial-gravity waves, and the background spiral wave. 
The development of the instability in the linear regime involves the growth of a broad spectrum of inertial modes, with growth rates on the order of the orbital time, and results in a nonlinear saturated state in which turbulent velocity perturbations are of a similar magnitude to those induced by the spiral wave. 
The turbulence induces angular momentum transport, and vertical mixing, at a rate that depends locally on the amplitude of the spiral wave (we obtain a stress parameter $\alpha \sim 5 \times 10^{-4}$ in our reference model). 
The instability is found to operate in a wide-range of disk models, including those with isothermal or adiabatic equations of state, and in viscous disks where the dimensionless kinematic viscosity $\nu\le 10^{-5}$. 
This robustness suggests that the instability will have applications to a broad range of astrophysical disk-related phenomena, including those in close binary systems, planets embedded in protoplanetary disks (including Jupiter in our own Solar System) and FU Orionis outburst models. 
Further work is required to determine the nature of the instability, and to evaluate its observational consequences, in physically more complete disk models than we have considered in this paper.
\end{abstract}


\section{INTRODUCTION}

Spiral density waves are excited in circumstellar disks by a variety of processes, and play an important role in the dynamical evolution and observational appearance of these systems. 
They arise as normal modes in self-gravitating disks \citep{lin64}, where they provide an efficient mechanism for the transport of mass and angular momentum through gravitational torques and an advective wave flux \citep{lyndenbell72,papaloizou91,laughlin94}. 
Spiral shocks can be excited by stellar companions in close binary systems, including cataclysmic variables, X-ray binaries and T Tauri binaries \citep{lin79, sawada86, artymowicz94}.
The excitation of nonlinear spiral waves by giant planets also plays an important role in structuring protoplanetary disks, and in driving the migration of embedded planets \citep{bryden99,kley99,nelson00}. 
Models for the origin of FU Orionis outbursts have been presented where spiral waves, originating in the gravitationally unstable outer regions of protostellar disks during the early infall phase, propagate into the inner disk regions and trigger the magnetorotational instability by heating and ionizing the disk gas there \citep{gammie99,armitage01,zhu10,bae14}. 
Recent work has also shown that the infall of low angular momentum material onto a protostellar disk during the early infall phase can also generate global spiral waves \citep{lesur15}.

In this paper, we show that spiral waves in circumstellar disks are unstable to the growth of a parametric instability that leads to the disk flow becoming turbulent. 
We came across this phenomenon while extending the two-dimensional hydrodynamic simulations of triggered FU Orionis outbursts in \cite{bae14} to three dimensions. 
The instability arises because pairs of inertial waves, or inertial-gravity waves, couple resonantly to the spiral wave, leading to the extraction of energy and angular momentum from it and the growth of a broad spectrum of inertial waves. 
Similar parametric instabilities, leading to the growth of inertial waves and the generation of small scale turbulence, have been reported to arise in numerous circumstances where disk fluid elements are subjected to periodic forcing. \cite{goodman93} and \cite{ryu94} showed that a parametric instability arises due to the elliptical distortion of a disk caused by an external orbiting companion. 
Warped disks have been shown to be subject to parametric instability \citep{gammie00, ogilvie13}, as have globally eccentric disks \citep{papaloizou05a, papaloizou05b, barker14}. 
In a study similar to the one that we present here, \cite{fromang07} used shearing box simulations and analytical calculations to show that axisymmetric, nonlinear sound waves traveling in disks are also subject to parametric instability.

In this paper, we examine the stability of propagating $m=2$ spiral waves in circumstellar disks, by carrying out three-dimensional global hydrodynamic simulations at high resolution. 
We consider a broad range of simplified disk models and input physics,  including both isothermal and adiabatic equations of state, viscous and inviscid disks, and vertically stratified and non-stratified density profiles. 
Simulations are computed with four independent numerical codes, and the outcomes are found to be in good agreement with each other. 
Our main result is that the spiral density waves are unstable to the growth of the aforementioned parametric instability, leading to the development of turbulence in the disk models. 
We refer to the instability as the spiral wave instability (SWI).
The SWI is found to be robust across the full range of models that we considered, suggesting that it will be influential in the dynamical evolution of disks that contain nonlinear spiral density waves of whatever origin.

This paper is organized as follows. 
In Section \ref{sec:background}, we discuss the theoretical background to the SWI to set the scene for the numerical simulations.
We describe our numerical methods in Section \ref{sec:method}, including the basic equations solved and the disk models examined.
In Section \ref{sec:cyl}, we present results from cylindrical disk models as a demonstration of the instability in the simplest global model with a non-stratified, globally isothermal initial setup.
Results from vertically stratified disk models with isothermal and adiabatic equations of state are discussed in Sections \ref{sec:iso} and \ref{sec:adia}, respectively.
We discuss the application of the SWI to various astrophysical disk phenomena, with a particular emphasis on protoplanetary disks, in Section \ref{sec:discussion}, and we provide concluding remarks in Section \ref{sec:conclusion}. 
A reader who is mainly interested in the physical manifestation of the instability, and its potential applications, can safely skip Sections \ref{sec:background} --  \ref{sec:cyl} and read from Section \ref{sec:iso} onwards.

\section{THEORETICAL BACKGROUND}
\label{sec:background}
Working in the context of a cylindrical coordinate system ($R$, $\phi$, $Z$), we consider the situation where a spiral density wave, with azimuthal mode number $m$, propagates inwards in a quasi-Keplerian disk. 
The disk is assumed to be isothermal in the vertical direction (i.e. temperature independent of $Z$). 
We consider models in which the thermodynamic response of the gas is adiabatic, such that $\gamma > 1$, and models where the disk gas responds isothermally such that $\gamma=1$. 
Under these conditions, the sound speed of the gas is independent of $Z$, and according to linear theory a spiral density wave that is launched from an inner Lindblad resonance (ILR) and has no vertical structure should propagate inwards maintaining its two-dimensional nature \citep{lin90}. 
Here, the absence of vertical structure simply means that perturbed quantities associated with the wave are independent of height.

A parametric instability may arise when an oscillator is subjected to a particular form of forcing at twice the natural frequency of the oscillator (e.g. a swinging pendulum with a moment of inertia that varies sinusoidally at twice the pendulum's mean natural frequency). 
In the context of astrophysical disks, parametric instabilities have been examined that arise from the excitation of pairs of inertial modes in disks that are elliptical due to tidal distortion by an external binary companion \citep{goodman93} or due to the propagation of axisymmetric, nonlinear sound waves \citep{fromang07}, where the two excited inertial waves have frequencies that satisfy the resonance condition $\omega_{\rm i, 1} + \omega_{\rm i, 2} = \omega_{\rm F}$, where the frequency of the forcing disturbance is $\omega_{\rm F}$. 
Viewed in a frame corotating with a fluid element in the disk, the frequency associated with a spiral wave that propagates interior to its inner Lindblad resonance is given by $\omega_{\rm s} = m (\Omega - \Omega_{\rm p})$, where $\Omega$ is the local orbital angular frequency and $\Omega_{\rm p}$ is the pattern speed of the spiral wave measured in the inertial frame. 
In the specific case of an external companion on a circular orbit, $\Omega_{\rm p}$ would be the angular frequency of the companion's orbit.

The WKBJ dispersion relation for local disturbances in a differentially rotating disk is given by \citep{goodman93}
\begin{equation}
\frac{\omega^2 /c_s^2}{\omega^2 - N^2} - \frac{k_ Z^2}{\omega^2 - N^2} - \frac{k_R^2}{\omega^2 - \kappa^2} = 0,
\label{eqn:dispersion1}
\end{equation}
where $\omega$ is the mode frequency, $\kappa$ is the epicyclic frequency defined as
\begin{equation}
\kappa^2 = \frac{1}{R^3} \frac{d}{dR} (R^2 \Omega)^2,
\label{eqn:kappa2}
\end{equation}
$N$ is the Brunt-V\"ais\"al\"a frequency defined as
\begin{equation}
N^2 = g \left(\frac{1}{\gamma P} \frac{dP}{dZ} - \frac{1}{\rho} \frac{d\rho}{dZ} \right),
\label{eqn:N2}
\end{equation}
$k_R$ and $k_Z$ are the wave numbers associated with the wave vector ${\bf k} = k_R {\bf {\hat e}}_R + k_Z {\bf {\hat e}}_Z$ and $c_s$ is the sound speed. In the high frequency limit, $\omega^2 \gg N^2$, $\kappa^2$, Equation~(\ref{eqn:dispersion1}) supports acoustic waves where $\omega^2 \simeq c_s^2 k^2$. 
In the low frequency limit, where $\omega^2 \ll c_s^2 k^2$, the dispersion relation becomes 
\begin{equation}
\omega_{\rm i}^2 = \kappa^2 \cos^2{\theta} + N^2 \sin^2{\theta},
\label{eqn:dispersion-inertialgravity}
\end{equation}
where the angle between the wave vector ${\bf k}$ and the $Z$-axis $\theta=\tan^{-1}(k_R/k_Z)$.
Equation~(\ref{eqn:dispersion-inertialgravity}) governs the behavior of inertial-gravity modes. 
These correspond to fluid elements undergoing epicyclic motions, with the fluid motions being confined to planes that are tilted by an angle $\theta$ with respect to the disk midplane, and which are perpendicular to the wave vector, ${\bf k}$. 
Coriolis and buoyancy forces provide the restoring forces in the horizontal and vertical directions, respectively. In the absence of vertical buoyancy, as in a disk with an isothermal response to perturbations ($N^2=0$ when $\gamma=1$), the dispersion relation simplifies to one that describes inertial waves
\begin{equation}
\omega_{\rm i}^2 = \kappa^2 \cos^2{\theta}.
\label{eqn:dispersion-inertial}  
\end{equation}
To simplify the discussion, we use the term \emph{inertial modes/waves} to denote inertial-gravity waves (supported by Coriolis and buoyancy forces) and inertial waves (supported only by the Coriolis force) from now on.

We might expect parametric instability to arise locally, through the coupling of a pair of inertial waves to an incoming spiral wave with doppler-shifted frequency $\omega_{\rm s}$, when the two inertial waves have frequencies that obey 
the relation $\omega_{\rm i, 1} + \omega_{\rm i, 2} = \omega_{\rm s}$. Working in the high wave number limit where $\omega_{\rm i, 1} \simeq  \omega_{\rm i, 2}$ \citep{fromang07}, we have
\begin{equation}
2 \omega_{\rm i} \simeq m ( \Omega - \Omega_{\rm p}).
\label{eqn:parametric1}
\end{equation}
For the specific case of an $m=2$ spiral wave, as considered in this paper, this becomes
\begin{equation}
\omega_{\rm i} = \Omega - \Omega_{\rm p}.
\label{eqn:parametric2}
\end{equation}
We see that the frequencies of inertial waves are independent of wave number, and simply depend on the orientation of the wave vector. 
For a disk that is finite in both radius and height, the wave numbers, $k_R$ and $k_Z$, and associated wave frequencies, will be discrete. 
Considering a situation where the radial and vertical wavelengths are given by $\lambda_R = H/n_R$ and $\lambda_Z=H/n_z$, respectively (where $H$ is the scale height), such that $k_R=2\pi n_R/H$, $k_Z=2\pi n_Z /H$ and ($n_R$, $n_Z$) are integers, then we have
\begin{equation}
\omega_{\rm i}^2 (n_R, n_Z) = \frac{n_Z^2}{n_R^2+n_Z^2} \kappa^2.
\label{eqn:omega_i_integer}
\end{equation}
It is obvious from Equation~(\ref{eqn:omega_i_integer}) that if we consider arbitrarily large $n_R$ and $n_Z$, then $\omega_{\rm i}$ can be matched with arbitrary precision to the value of $\Omega - \Omega_{\rm p}$, where $0 < \Omega_{\rm p} < \Omega$. 
In other words, inertial waves have a dense spectrum, and the resonance condition can always be matched to high precision for large enough $k$. \citet{fromang07}, considering a disk with an isothermal equation of state, comment on the fact that the density of inertial modes is greater in a disk with vertical density stratification than in a non-stratified disk because of the larger range of length scales. 
Here, we also note that an adiabatic disk with $N^2 >0$ also increases the density of inertial modes, because $N^2$ scans a range of frequencies as a function of height, and so increases the probability that the resonant condition for the parametric instability can be met at any given radius (i.e. one combination of $n_R$ and $n_Z$ may satisfy the resonance condition near the midplane where $N^2 \simeq 0$, and another combination may satisfy the resonance condition away from the midplane where $N^2 > 0$).

\subsection{Wavelengths of parametrically excited inertial modes}
\label{sec:mode-length}
Obtaining a full and detailed understanding of which specific pairs of inertial modes will couple to an incoming spiral wave by satisfying the resonance condition $\omega_{\rm i,1} + \omega_{\rm i,2} = \omega_{\rm s}$ would require determination of the radial and vertical eigenmode structure, and associated eigenfrequencies, for the global disk models that we consider. 
This goes beyond the scope of this paper (but see \cite{barker15} for a study of a similar problem in the context of the vertical shear instability). 
Instead, we use simple arguments, based on a local WKBJ picture, to consider the approximate wave numbers associated with unstable inertial modes to help interpret and understand the simulations presented in later sections.

The WKBJ dispersion relation for a spiral wave with azimuthal mode number $m$ (in the absence of self-gravity) is given by (assuming $k_Z=0$)
\begin{equation}
m^2 (\Omega - \Omega_{\rm p})^2 =\kappa^2 + c_s^2 k_{R,{\rm s}}^2,
\label{eqn:spiral-dispersion}
\end{equation}
giving a radial wave number interior to the ILR
\begin{equation}
k_{R, {\rm s}}^2 = \frac{m^2 (\Omega - \Omega_{\rm p})^2 - \kappa^2}{c_s^2}.
\label{eqn:k-spiral}
\end{equation}
Inertial modes that are excited by the spiral wave must occur on radial length scales similar to or smaller than the incoming wavelength, so for simplicity we write the wave number of the excited inertial waves (assuming they have very similar spatial structure and frequencies, appropriate to the high wave number limit) $k_{R, {\rm i} }= n k_{R, {\rm s}}$, where $n$ is an integer. 
Given the radial wave number and the frequency of the spiral wave (in the local fluid frame), we can estimate the vertical wave number of the excited inertial waves using the dispersion relation from Equation~(\ref{eqn:dispersion1}), written using the modified notation
\begin{equation}
\frac{k_{Z, {\rm i}}^2}{\omega_{\rm i}^2-N^2} + \frac{k_{R, {\rm i}}^2}{\omega_{\rm i}^2 - \kappa^2} - \frac{\omega_{\rm i}^2/c_s^2}{\omega_{\rm i}^2 - N^2} = 0,
\label{eqn:dispersion3}
\end{equation}
giving
\begin{equation}
k_{Z,{\rm i}}^2 = - \frac{n^2 k_{R,{\rm s}}^2}{\omega_{\rm i}^2 - \kappa^2} (\omega_{\rm i}^2 - N^2) + {\omega_{\rm i}^2 \over c_s^2}.
\label{eqn:kZi}
\end{equation}
Substituting the resonance condition $\omega_{\rm i}=\omega_{\rm s}/2$ and $k_{R, {\rm s}}^2 = (\omega_{\rm s}^2 - \kappa^2)/c_s^2$ into Equation~(\ref{eqn:kZi}) gives
\begin{equation}
k_{Z,{\rm i}}^2 = \frac{(\omega_{\rm s}/2)^2}{c_s^2} - n^2 \frac{\omega_{\rm s}^2- \kappa^2}{c_s^2} \frac{(\omega_{\rm s}/2)^2 - N^2}{(\omega_{\rm s}/2)^2 - \kappa^2}.
\label{eqn:kZi2}
\end{equation}
First, we consider the behavior of $k_{Z,{\rm i}}$ for disks in which $N^2=0$
\begin{equation}
k_{Z,{\rm i}}^2 = \frac{(\omega_{\rm s}/2)^2}{c_s^2} \left(1 - n^2 \frac{\omega_{\rm s}^2-\kappa^2}{(\omega_{\rm s}/2)^2 - \kappa^2} \right).
\label{eqn:kZi3}
\end{equation}
For $k_{Z,{\rm i}}$ to be real, such that inertial waves can be excited, we require that $\omega_{\rm s}^2 > \kappa^2$ and $(\omega_{\rm s}/2)^2 < \kappa^2$, and these conditions are both satisfied at all radii interior to the inner Lindblad resonance of an $m=2$ spiral wave in a Keplerian disk. (These conditions, however, are not satisfied everywhere for a spiral wave that propagates outwards from an outer Lindblad resonance, as we discuss later in Section~\ref{sec:discussion}.)
Considering the influence of buoyancy in Equation~(\ref{eqn:kZi2}), we see that $k_{Z,{\rm i}}$ can become imaginary when $(\omega_{\rm s}/2)^2 < N^2$, indicating that inertial modes will not be excited at certain locations above the midplane where the buoyancy frequency is large. 
Taking the large $n$ limit, the transition occurs at $(\omega_{\rm s}/2)^2 = N^2$. 
Given that the resonance condition for the excited inertial waves is $\omega_{\rm i} = (\omega_{\rm s}/2)$, this indicates that inertial modes cannot be excited where $\omega_{\rm i} < N$. 
Later in the paper, we use Equation~(\ref{eqn:kZi2}) to construct maps of stable and unstable regions of our disks for comparison with the simulation results, and to consider the influence of numerical resolution on our ability to resolve unstable inertial waves that arise on small scales.

As we have emphasized, there are a number of caveats that should be considered before applying the above arguments to the simulation results. 
The most obvious is the fact that a local WKBJ analysis, that applies to modes where $k_Z H \gg 1$ and $k_R H \gg 1$, is not strictly applicable to modes with $k_R \sim 1/H$, as may be excited by spiral waves in a global disk model. 
Furthermore, for simplicity we have assumed that there is an integer relation between the wavelengths of the excited inertial waves and the incoming spiral modes, and this is not precisely the situation that should arise because the linear inertial modes supported in the global disk model should be quantized to fit within the radial boundaries of the disk, and hence are not expected to have an integer relation with the spiral waves. 
A full analysis of the modes that can be excited should take account of the full structure of the eigenfunctions in both radial and vertical directions, and their frequencies, which represents a significant computational problem \cite[e.g.][]{barker15}, and goes beyond the scope of this paper. 
It is noteworthy, however, that \cite{lubow93} compute the vertical structure of inertial-gravity waves in vertically isothermal disks that support buoyancy forces, and demonstrate that inward traveling waves have their wave energy increasingly confined towards the midplane as they enter disk radii where large fractions of the disk vertical domain have $N^2 > \omega_{\rm i}^2$, since this defines the region of space where the waves can (or cannot) propagate. 
Similar forbidden zones arise in the above WKBJ analysis for essentially the same reason, as we have described above.

\subsection{Nonlinear wave propagation in vertically stratified disks}
\label{sec:nonlinear}

In linear theory, a two-dimensional spiral wave without any vertical structure, launched at an inner Lindblad resonance in a vertically isothermal disk with sound speed independent of $Z$, should propagate towards the star with wave fronts remaining perpendicular to the midplane. 
As described in the next section,  the underlying axisymmetric disk model will normally have midplane density varying according to $\rho(R,0) = \rho(R_0,0) (R/R_0)^{p}$, with $p<0$, and variation with height scaling as $\rho(R,Z) \sim \rho(R,0) \exp{(-Z^2/(2H^2))}$, where $H$ is the density scale height. 
For such a model, the wave front associated with an inward propagating wave will see an increasing midplane density as it moves in, but at high altitudes the wave front will see a sharply decreasing density. 
We might then expect that nonlinear effects experienced by the wave to be greater at high altitudes than near the midplane. 
These effects include distortion of the wave form and an enhanced propagation speed due to advection of the wave by the perturbed radial velocity. 
We expect this will result in curvature of the spiral wave fronts as these waves propagate towards the star. 
An important consequence of this is that vertical hydrostatic equilibrium will not be maintained at the wave fronts, resulting in the generation of vertical motions as the wave moves inwards, and an oscillating non-axisymmetric corrugation of the disk surface.

\section{NUMERICAL METHODS}
\label{sec:method}

\subsection{Basic Equations}

We solve the hydrodynamic continuity, momentum, and internal energy equations:
\be\label{eqn:mass}
{\partial \rho \over \partial t} + \nabla \cdot (\rho v) = 0,
\en
\be\label{eqn:momentum}
\rho \left( {\partial v \over \partial t} + v \cdot \nabla v \right) = - \nabla P - \rho \nabla (\Phi_* + \Phi_p)+ \nabla \cdot \Pi,
\en
\be\label{eqn:energy}
{\partial e \over \partial t} + \nabla \cdot (ev) = - P \nabla \cdot v,
\en
where $\rho$ is the mass density, $v$ is the velocity, $P$ is the pressure, $\Phi_*$ is the gravitational potential due to the central object, $\Phi_p$ is the spiral potential (see Section \ref{sec:potmodel}), $\Pi$ is the viscous stress tensor, and $e$ is the internal energy per unit volume.

We make use of both cylindrical $(R, \phi, Z)$ and spherical $(r, \theta, \phi)$ coordinates.
We consider cylindrical disk models where we neglect the vertical component of gravity from the central object, so the gravitational potential is  $\Phi_* = GM_*/R$, where $G$ is the gravitational constant, $M_*$ is the mass of the central object and $R$ is the cylindrical radius. 
These models are computed using cylindrical coordinates.
The purpose of using non-stratified, cylindrical models is to illustrate the nature of the instability as clearly as possible, by removing/minimizing complications that may arise from spherical geometry, concave wave structure (see Section \ref{sec:ref}), etc.
Furthermore, adopting a cylindrical disk model allows us to run models with periodic boundary conditions in the vertical direction, and hence allows us to demonstrate that the existence of the instability is independent of the chosen boundary conditions.

We also consider models where the density is vertically stratified. 
The gravitational potential is then $\Phi_* = GM_*/r$, where $r=\sqrt{R^2+Z^2}$ is the spherical radius. 
These models are computed using spherical coordinates.

We make use of two forms of equation of state.
In the isothermal calculations, we use an isothermal equation of state $P = \rho c_s^2$, where $c_s$ is the isothermal sound speed, and do not solve the internal energy equation.
In the adiabatic calculations, we relate the gas pressure and the internal energy through $P=(\gamma-1)e$, where $\gamma$ is the adiabatic index, and solve the internal energy equation.
We do not include the effects of cooling in the internal energy equation. Instead, we explore the effects of cooling by changing the adiabatic index value. The idea is that a disk with very rapid cooling acts as if it is isothermal, and hence has an effective $\gamma \sim1$, whereas a disk with very slow cooling acts as if it is adiabatic, and thus its effective $\gamma$ is equal to the actual ratio of specific heats in the disk, here taken to be $\gamma=1.4$. 
Intermediate values of the effective $\gamma$ correspond to intermediate cooling rates. We take this approach because determining the Brunt-V\"ais\"al\"a frequency is not trivial in the presence of cooling, making comparison between numerical results and theoretical predictions difficult.

We include physical and artificial viscosity \citep{stone92}. 
\citet{lyra16} has recently considered the local generation of entropy through strong spiral shocks induced by massive planets, that could in turn drive local convective motions.
We have confirmed, however, that the shock dissipation heating is negligible in our adiabatic calculations because the spiral waves considered in this work are much weaker.

\begin{deluxetable*}{lccccccccccc}
\tablecolumns{15}
\tabletypesize{\small}
\tablecaption{Model Parameters \label{tab:parameters}}
\tablewidth{0pt}
\tablehead{
\colhead{Run Label} & 
\colhead{Code} & 
\colhead{Numerical Resolution} & 
\colhead{$\mathcal{A}$} & 
\colhead{Kinematic} & 
\colhead{$\gamma$} & \\
\colhead{} & 
\colhead{} & 
\colhead{($N_R$ or $N_r \times N_\phi \times N_z$ or $N_\theta$ )} & 
\colhead{} & 
\colhead{Viscosity $\nu$} &
\colhead{} & 
 }
\startdata
CYL-F & FARGO3D & $512\times128\times128$  & $5.0\times10^{-4}$ & 0 & 1.0 \\
CYL-N & NIRVANA & $3200\times128\times128$  & $5.0\times10^{-4}$ & 0 & 1.0 \\
CYL-P & PLUTO &$512\times128\times128$ & $5.0\times10^{-4}$  & 0 & 1.0 \\
CYL-I & INABA3D & $3200\times128\times128$ & $5.0\times10^{-4}$ & 0 & 1.0 \\
\hline
R512 & FARGO3D & $512\times128\times128$ & $5.0\times10^{-4}$ & 0 & 1.0 \\
\hline
R128 & FARGO3D & $128\times32\times32$ & $5.0\times10^{-4}$ & 0  & 1.0 \\
R256 & FARGO3D & $256\times64\times64$ & $5.0\times10^{-4}$ & 0 & 1.0 \\
R768 & FARGO3D & $768\times192\times192$ & $5.0\times10^{-4}$ & 0 & 1.0 \\
\hline
AMP0.625 & FARGO3D & $512\times128\times128$ & $6.25\times10^{-5}$ & 0 & 1.0 \\
AMP1.25 & FARGO3D & $512\times128\times128$ & $1.25\times10^{-4}$ & 0 & 1.0 \\
AMP2.5 & FARGO3D & $512\times128\times128$ & $2.5\times10^{-4}$ & 0 & 1.0 \\
AMP10 & FARGO3D & $512\times128\times128$ & $1.0\times10^{-3}$ & 0 & 1.0 \\
\hline
V6 & FARGO3D & $512\times128\times128$ & $5.0\times10^{-4}$ & $10^{-6}$ & 1.0 \\
V5 & FARGO3D & $512\times128\times128$ & $5.0\times10^{-4}$ & $10^{-5}$ & 1.0 \\
V4 & FARGO3D & $512\times128\times128$ & $5.0\times10^{-4}$ & $10^{-4}$ & 1.0 \\
\hline
GAM01 & FARGO3D & $512\times128\times128$ & $6.25\times10^{-5}$ & 0 & 1.01 \\
GAM05 & FARGO3D & $512\times128\times128$ & $6.25\times10^{-5}$ & 0 & 1.05 \\
GAM1 & FARGO3D & $512\times128\times128$ & $6.25\times10^{-5}$ & 0 & 1.1 \\
GAM2 & FARGO3D & $512\times128\times128$ & $6.25\times10^{-5}$ & 0 & 1.2 \\
GAM3 & FARGO3D & $512\times128\times128$ & $6.25\times10^{-5}$ & 0 & 1.3 \\
GAM4 & FARGO3D & $512\times128\times128$ & $6.25\times10^{-5}$ & 0 & 1.4
\enddata
\end{deluxetable*}

\subsection{Disk Models}
\label{sec:model}

We begin with an initial radial power-law temperature distribution in the disk that is independent of height
\be
\label{eqn:temp}
T(R) = T_0 \left( {R \over R_0} \right)^q,
\en 
where $T_0$ is the temperature at $R_0=1$.
In all models presented, $T_0$ is chosen such that the ratio of disk scale height to radius at $R_0$ is $H_0/R_0 = 0.05$.
The isothermal sound speed is related to the temperature by $c_s^2=\mathcal{R} T/\mu$, where $\mathcal{R}$ is the gas constant and $\mu$ is the mean molecular weight.
Thus, Equation (\ref{eqn:temp}) corresponds to the radial sound speed distribution given by
\be
c_s(R) = {H_0 \over R_0} \left( {R \over R_0} \right)^{q/2}.
\en 

For the main simulation set we assume $q=0$, so that the disk has the same temperature everywhere. This temperature structure might not be very realistic, especially when the simulation domain extends more than an order of magnitude in radius.
On the other hand, one can safely ignore the vertical shear instability in these models, because the instability requires a non-zero vertical gradient in the disk angular velocity which arises when there is a radial temperature gradient \citep{nelson13}.
In the Appendix, we present models with a radial temperature gradient ($q=-1$).
In those models, we find that the spiral wave instability and the vertical shear instability coexist in the absence of kinematic viscosity, and thus identifying intrinsic features arising from the SWI is challenging.
The vertical shear instability can be suppressed with a non-zero kinematic viscosity, but adding kinematic viscosity not only suppresses the vertical shear instability but also damps the growth of small scale unstable modes excited by the SWI.
To summarize, by assuming the globally isothermal temperature structure, we aim to illustrate the nature of the SWI as clearly as possible, in the absence of any other hydrodynamic instabilities.

The initial density and azimuthal velocity profiles are constructed to satisfy hydrostatic equilibrium \citep[e.g.][]{nelson13}: 
\be
\label{eqn:init_den}
\rho(R,Z)   =  \rho_0 \left({R \over R_0}\right)^{p} \exp\left({GM_* \over c_s^2} \left[{1 \over \sqrt{R^2+Z^2}} - {1 \over R}\right]\right)
\en
and
\be
\label{eqn:init_vel}
v_\phi(R,Z)  =\left[ (1+q){GM_* \over R} + \left( p + q \right) {c_s^2} - q{GM_* \over \sqrt{R^2+Z^2}} \right] ^{1/2}.
\en
The midplane density at $R=R_0$, $\rho_0$, is chosen such that the total disk mass is $10~\%$ of the central object's mass.
When vertical stratification of the disk is ignored in the cylindrical disk models, the above equations simplify to: 
\be
\rho(R,Z)  =  \rho_0 \left({R \over R_0}\right)^{p} 
\en
and
\be
v_\phi(R,Z)  =  \left[ {GM_* \over R} + \left( p + q \right) {c_s^2} \right] ^{1/2}.
\en
We fix the radial power-law index of the midplane gas density to $p=-1.5$ for all models.
The initial radial and vertical/meridional velocity are set to zero, but uniformly distributed random perturbations are added as white noise, at the level of $10^{-6}~c_s$, to the initial vertical/meridional velocity in order to seed the instability.
We have tested the effects of changing the amplitudes of the random perturbations, and found that the growth rate and the saturation level of the instability are not sensitive to the initial noise level.

Our simulation domain extends from $0.5$ to 10 in radius and from 0 to $2\pi$ in azimuth.
In cylindrical models, the vertical domain extends from $Z=-0.1$ to 0.1.
In spherical models, the meridional domain covers $\pm 4 H_0/R_0$ above and below the midplane. The model parameters are summarized in Table \ref{tab:parameters}.

\subsection{Spiral Potential}
\label{sec:potmodel}

\begin{figure*}
\centering
\epsscale{1.15}
\plotone{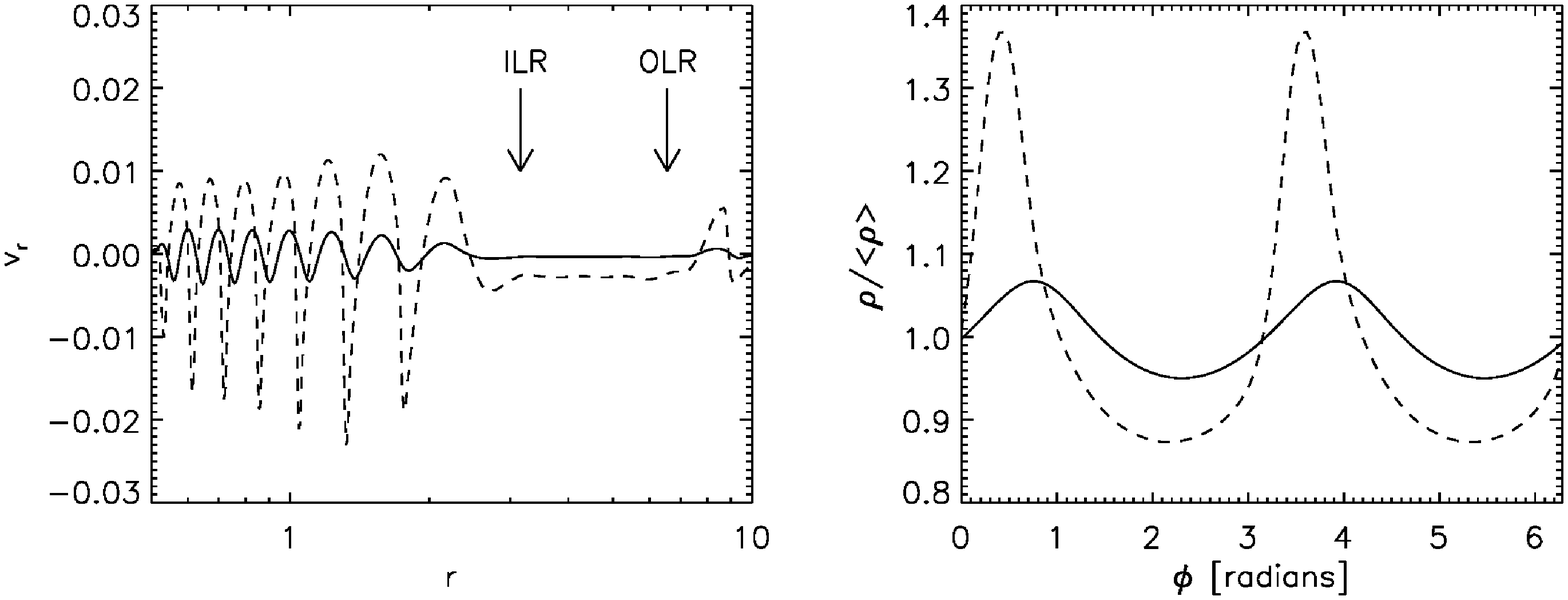}
\caption{(Left) The radial velocity $v_r$ in the disk midplane induced by the spiral potential with amplitudes of (solid) $\mathcal{A}=6.25\times10^{-5}$ from model AMP0.625 and (dashed) $\mathcal{A}=5\times10^{-4}$ from model R512. The arrows indicate the location of the inner and outer Lindblad resonances (ILR and OLR). (Right) The azimuthal distribution of the midplane density at $r=0.6$, normalized by the azimuthal average. As in the left panel, the solid and dashed curves are when $\mathcal{A}=6.25\times10^{-5}$ and $\mathcal{A}=5\times10^{-4}$, respectively. The data are taken after the spiral waves are well established over the entire disk, but before the spiral wave instability perturbs the wave structures.}
\label{fig:waves}
\end{figure*}

In order to excite spiral waves, we adopt the following potential $\Phi_p$:
\be
\label{eqn:potmodel}
\Phi_p(R, \phi, t) = \mathcal{A} \cos[m(\phi - \Omega_p t)] e^{-(R-R_p)^2/\sigma_p^2},
\en
where $\mathcal{A}$ is the amplitude, assumed to be constant over time, $m$ is the azimuthal mode number, $\Omega_p$ is the pattern speed, $R_p$ is the radius about which the potential is centered, and $\sigma_p$ is the radial width of the potential.
We adopt the values $R_p = 5$ and $\sigma_p=1$, and assume that the pattern speed is the local Keplerian frequency at its central position: $\Omega_p = (GM_*/R_p^3)^{1/2}$.
In this work, we only focus on spiral potentials with $m=2$.

Our fiducial model assumes potential amplitude of $\mathcal{A}=5.0\times10^{-4}$.
With this amplitude we intend to produce velocity perturbations at $R  = 1$ that are about $\sim 20 \%$ of the local sound speed\footnote{As we mentioned in the introduction, the spiral wave instability was found while studying spiral wave propagation as part of a study of accretion outbursts in FU Orionis systems. 
This spiral amplitude produces velocity and density fluctuations in the inner disk similar to the ones produced by the gravitationally unstable outer disk, with which an accretion outburst was triggered (see Figure 5b of \citealt{bae14}).}.
The effect of varying the spiral potential strength on the spiral wave instability is investigated in Section \ref{sec:pstrength}.

Prior to discussing the results of simulations in any detail, we briefly illustrate the effects of changing the spiral wave amplitude on the wave forms and propagation speeds of the waves in the midplane of a stratified disk. 
In Figure \ref{fig:waves}, we present the radial velocity profiles in the disk midplane of a stratified disk model, driven by two  spiral potential amplitudes of $\mathcal{A} = 5.0\times10^{-4}$ and $\mathcal{A} = 6.25\times10^{-5}$.
As shown, the overall structure of the spiral waves driven in the numerical simulations agrees well with theoretical expectations: from the WKBJ dispersion relation for a non self-gravitating disk one expects to see waves propagating interior and exterior to the inner and outer Lindblad resonances, with the waves being evanescent in the region between the resonances  \citep{lin64,goldreich78,goldreich79}.
The wave with the small potential amplitude $\mathcal{A}=6.25\times10^{-5}$ is nearly sinusoidal. In the inner disk, the velocity amplitude is about a few per cent of the sound speed and is nearly constant over radius. When the spiral potential amplitude is increased to $\mathcal{A}=5\times10^{-4}$, on the other hand, the waves are rather strong, generating sharp wave fronts that have radial velocities up to about $40~\%$ of the sound speed.
At $r=0.6$, the two spiral potential amplitudes generate relative density enhancements of $\sim 7 \%$ ($\mathcal{A}=6.25\times10^{-5}$) and $\sim 37 \%$ ($\mathcal{A}=5\times10^{-4}$) above the azimuthal average.
In addition to these differences in perturbed velocities and densities, we also note that the propagation speed of the higher amplitude wave is slightly faster than that of the lower amplitude wave. 
This is a nonlinear effect arising from the fact that the magnitude of the perturbed velocity contributes to the wave speed through the advection term in the momentum equation. 
The magnitude of this effect as a function of height in the disk plays an important role in determining the shape of the spiral wave fronts in the stratified disk models, as discussed in Section~\ref{sec:background} and demonstrated in Section~\ref{sec:iso}.

\subsection{Boundary Conditions}

One needs to take particular care with the boundary conditions when simulating an instability, in order to ensure that what is being observed is not an artifact generated by the boundaries.
The requirement that models with vertical stratification are able to maintain hydrostatic equilibrium has forced us to pay particular attention to the meridional boundary conditions.
The radial boundary condition is chosen to have a zero gradient for all variables in all models, 
and a wave damping zone is implemented \citep{devalborro06} in the intervals $R=[0.5,0.6]$ and $R=[9,10]$.  
We choose one local orbital time for the wave damping timescale.
As the simulation domain covers $2\pi$ in azimuth, periodic boundary conditions are used in azimuth.

Vertical periodic boundary conditions are used in models where we neglect vertical stratification.
In the stratified models, we use an outflow meridional boundary condition such that all velocity components in the ghost zones have the same values as the last active zones, but the meridional velocity is set to 0 if directing toward the disk midplane. In adiabatic calculations, the temperature in the ghost zones is set to have the same value as in the last active zones. The density in the ghost zone is then obtained by solving the hydrostatic equilibrium in the meridional direction:
\be
\label{eqn:hsebc}
{1 \over \rho} {\partial \over \partial\theta}{(\rho c_s^2)} = {v_\phi^2 \over \tan\theta}.
\en 

In order to investigate the influence of the meridional boundaries in the stratified models, we tested various boundary conditions including the zero-gradient boundary condition, the standard outflow boundary condition, the reflecting boundary condition, and the wave-damping boundary conditions.
We find that the choice of the meridional boundary condition does not affect the triggering of the parametric instability. 
For the Godunov codes, however, we find that enforcing hydrostatic equilibrium in the meridional ghost zones using Equation (\ref{eqn:hsebc}) maintains the low-density surface region more stably than with other boundary conditions, allowing us to better observe the early development of the instability.

\subsection{Codes}
\label{sec:codes}

We use four independent grid-based codes: FARGO3D, NIRVANA, PLUTO, and INABA3D.
These include two finite difference codes and two Godunov codes, to ensure the robustness of the results.

FARGO3D is a finite difference code developed with special emphasis on disk simulations \citep{benitez16}.
An orbital advection algorithm is implemented as in its two-dimensional predecessor FARGO (Fast Advection in Rotating Gaseous Objects; \citealt{masset00}) code.
We have tested with and without the FARGO algorithm and found that the result is not dependent on the use of the FARGO algorithm.

Similar to FARGO3D, NIRVANA uses an algorithm very similar to that used in the ZEUS code to solve the equations of ideal MHD \citep{ziegler97, stone92}. This scheme uses operator splitting, dividing the governing equations into source and transport terms. Advection is performed using the second-order monotonic transport scheme \citep{vanleer77}.

PLUTO is a general-purpose Godunov code \citep{mignone07}.
We employ the piece-wise parabolic reconstruction and third-order Runge-Kutta time integration.
We note that lower-order schemes, e.g. the piece-wise linear reconstruction and second-order Runge-Kutta time integration, do not produce a converged Riemann solution for the problem considered in this work.
The FARGO orbital advection module was enabled for the calculations with PLUTO.

INABA3D is a finite volume code using an unsplit MUSCL Hancock scheme, which is a second order method in time and space. 
The physical values on the faces of each cells are calculated through an exact Riemann solver.
The code is based on the 2D code described in \citet{inaba05}, and the third dimension has been implemented as in \citet{richard13}.

As shown in Table~\ref{tab:parameters}, most simulations in this paper were computed using the FARGO3D code. We note that FARGO3D and PLUTO use a logarithmically spaced radial grid, whereas NIRVANA and INABA3D use a uniform mesh. 
The innermost grid cell size in the FARGO3D and PLUTO runs was $\Delta r=0.002934$, and this value was used for the uniform meshes in the NIRVANA and INABA3D runs.

\subsection{Code Units}

We choose code units such that $GM_* = R_0 = \Omega (R=R_0) = 1$.
Since we do not include cooling, all the calculations are scalable to any system of interest. 
In the following sections we will use the orbital time at $R=1$ as the time unit.
We will denote this quantity as $1~T_{\rm orb}$.
One orbital time at the spiral potential center $R=R_p$, $T_p$, corresponds to $11.2~T_{\rm orb}$.

\begin{figure}
\centering
\epsscale{1.2}
\plotone{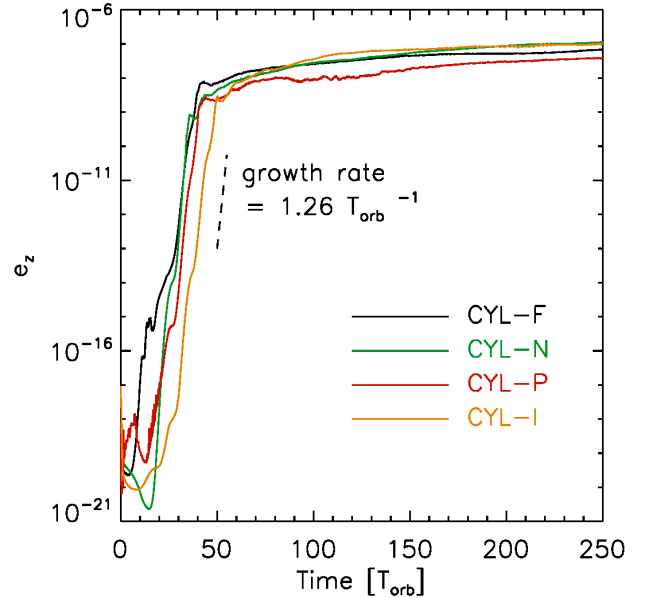}
\caption{Time evolution of the integrated vertical kinetic energy $e_Z$ for non-stratified models with different codes (CYL-F, CYL-N, CYL-P, CYL-I). The black dashed line corresponds to the growth rate of $1.26~T_{\rm orb}^{-1}$.}
\label{fig:kinz_cyl}
\end{figure}

\subsection{Diagnostics}

In order to examine the growth and evolution of the spiral wave instability we calculate the volume-integrated vertical kinetic energy, $e_Z$, in non-stratified disk models (as these are computed using cylindrical coordinates) and the volume-integrated meridional kinetic energy, $e_\theta$, in the stratified disk models (as these use spherical polar coordinates): 
\be
e_{Z,\theta} = {1 \over 2} \int_{V}{\rho v_{Z,\theta}^2}~dV.
\en
The volume integration is performed in between $R=0.6$ and $R=1.0$, in order to capture growth in a local region of the disk to better measure exponential growth rates.

\begin{figure*}
\centering
\epsscale{1.2}
\plotone{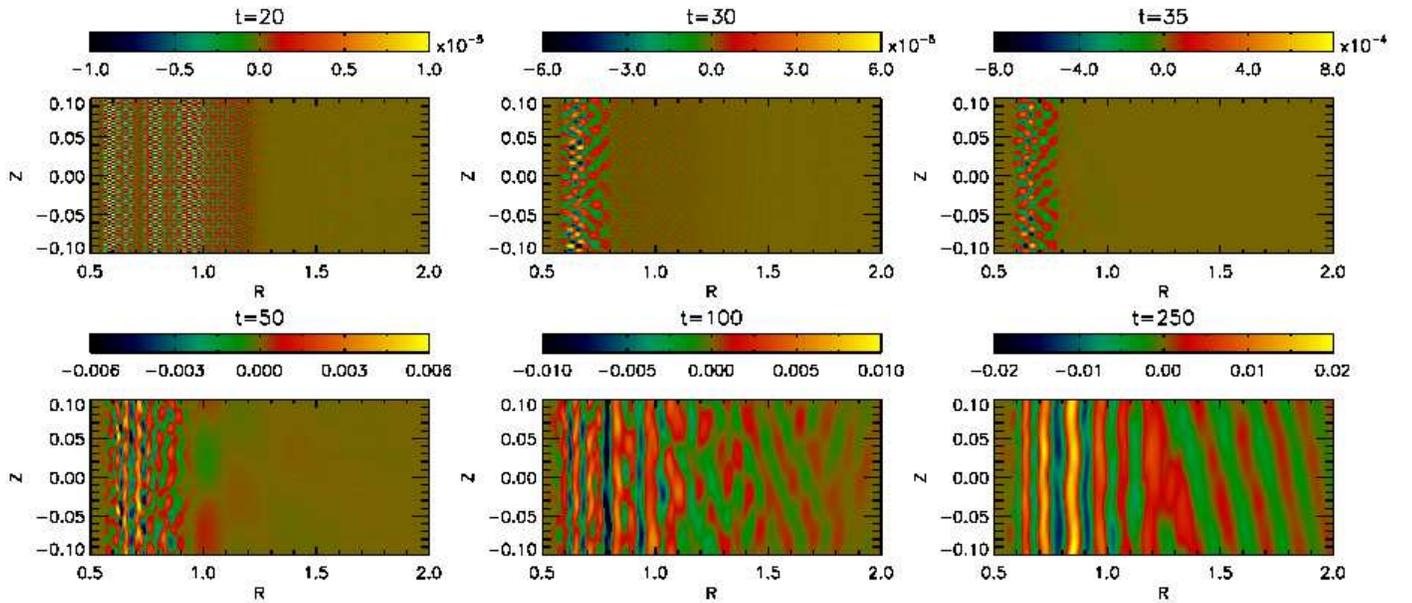}
\caption{Distributions of the vertical velocity in a $R-Z$ plane ($\phi=0$) interior to $R=2$ at various times for model CYL-F. The upper panels show the linear growth phase of the SWI. The instability saturates when the perturbed vertical velocity of the unstable modes is comparable to the radial velocity of the imposed spiral waves.}
\label{fig:cylz-f}
\end{figure*}

\begin{figure*}
\centering
\epsscale{1.2}
\plotone{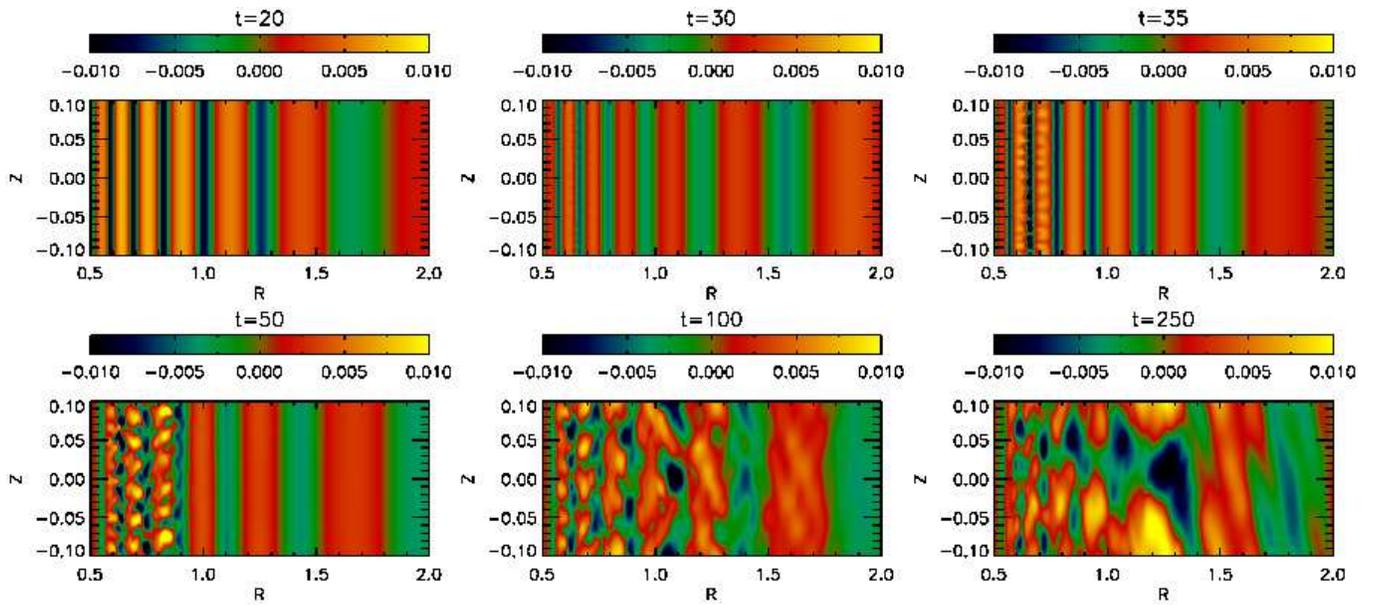}
\caption{Same as Figure \ref{fig:cylz-f}, but for the radial velocity.}
\label{fig:cylr-f}
\end{figure*}

\section{NON-STRATIFIED DISK MODELS}
\label{sec:cyl}

We start discussion of the simulations by describing results of the non-stratified disk models.
In Figure \ref{fig:kinz_cyl}, we present the time evolution of the integrated vertical kinetic energy $e_Z$, where we see that this quantity exponentially increases at early times ($t \lesssim 40$) before saturating to a quasi-steady value at late times, indicating the presence of an instability.

\begin{figure*}
\centering
\epsscale{1.15}
\plotone{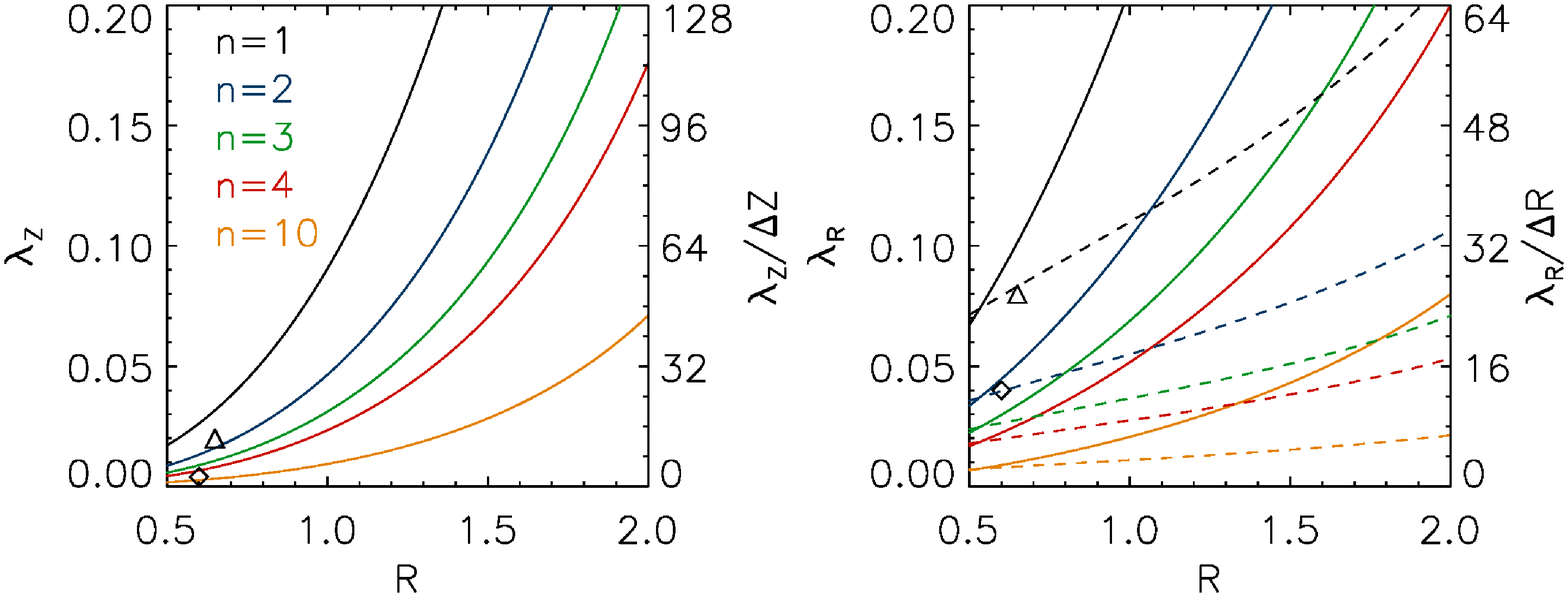}
\caption{The predicted (left) vertical and (right) radial wavelengths of the unstable modes $\lambda_Z$ and $\lambda_R$ as a function of radius, calculated from the relation given in Equation (\ref{eqn:kZi2}) and Equation (\ref{eqn:k-spiral}) with $k_{R, {\rm i} }= n k_{R, {\rm s}}$, where $n$ is an integer,  and $k_{R,{\rm s}}$ and $k_{R, {\rm i}}$ are the wave numbers of the spiral wave and the excited inertial modes, respectively.  The over-plotted symbols indicate the measured vertical and radial wavelengths of the unstable inertial modes at the innermost spiral arm location at $t=20$ ($R \sim 0.6$; diamonds) and 30 ($R \sim 0.65$; triangles). In the right panel, the solid curves present the radial wavelengths and the dashed curves show the number of radial grid cells in one inertial mode wavelength ($\lambda_R/\Delta R$).}
\label{fig:dr_cyl}
\end{figure*}

In Figures \ref{fig:cylz-f} and \ref{fig:cylr-f}, we display the two-dimensional distributions of the vertical and radial velocities at some selected times. 
Since vertical density stratification of the disk is neglected, and the sound speed is independent of height, the spiral waves propagate purely radially before the SWI sets in, maintaining wave fronts that are perpendicular to the disk midplane, as seen in the first two panels of Figure \ref{fig:cylr-f}. 
Hence, the spiral waves in themselves are not expected to generate any vertical motion in the disk. We see that the inertial modes grow fastest in the inner disk regions, as their frequencies and growth rates are larger there, and the instability spreads to larger radii as time progresses. At $t=20$, the inertial modes that are excited by the spiral wave form a checkerboard pattern in the vertical velocities, as seen in Figure \ref{fig:cylz-f}. 
Close visual inspection of the plot, combined with a Fourier analysis, shows that in the disk inner regions around $R=0.6$, the vertical wavelength associated with the excited inertial modes $\lambda_Z \simeq 0.004$ and the radial wavelength $\lambda_R \simeq 0.04$.
These wavelengths correspond to approximately 3 and 12 grid cell spacings in the vertical and radial directions, respectively, indicating that the fastest growing modes have the shortest wavelengths that can be represented on the computational grid, presumably limited by the vertical resolution in this case.
This is in agreement with \cite{fromang07}, who showed that short wavelength inertial waves excited by axisymmetric, nonlinear sound waves have higher growth rates than longer wavelength modes. 

Figure \ref{fig:cylz-f} shows that at slightly later times, $t=30$, the length scales associated with the most prominent perturbations have increased substantially, presumably indicating that these longer wavelength modes grow on longer time scales, but may also contain more energy than the shorter wavelength modes seen to grow earlier. 
This latter point may, however, be affected by numerical diffusion damping the smallest scale modes, so some caution is required when interpreting this result. Inspection of the figure, and Fourier analysis, indicate that these modes have wavelengths $\lambda_Z \simeq 0.02$ and $\lambda_R \simeq 0.08$.
We have plotted these and the earlier wavelength values in Figure~\ref{fig:dr_cyl}, which shows the vertical and radial wavelengths of the inertial modes that are predicted to be excited by the spiral wave using the simple local analysis presented in Section~\ref{sec:background}. 
Noting the caveats associated with the analysis discussed earlier in the paper, we see that the early growing modes have radial wavelengths that are $\sim 1/3$ that of the incoming spiral wave, and the later growing modes have radial wavelengths very similar to the spiral wave. 
The numbers of grid cells covered by the different wavelengths are also indicated in this figure, demonstrating that the earlier growing modes occur on length scales close to the grid scale, whereas the later growing modes are more comfortably represented on the grid.

\begin{figure}
\centering
\epsscale{1.2}
\plotone{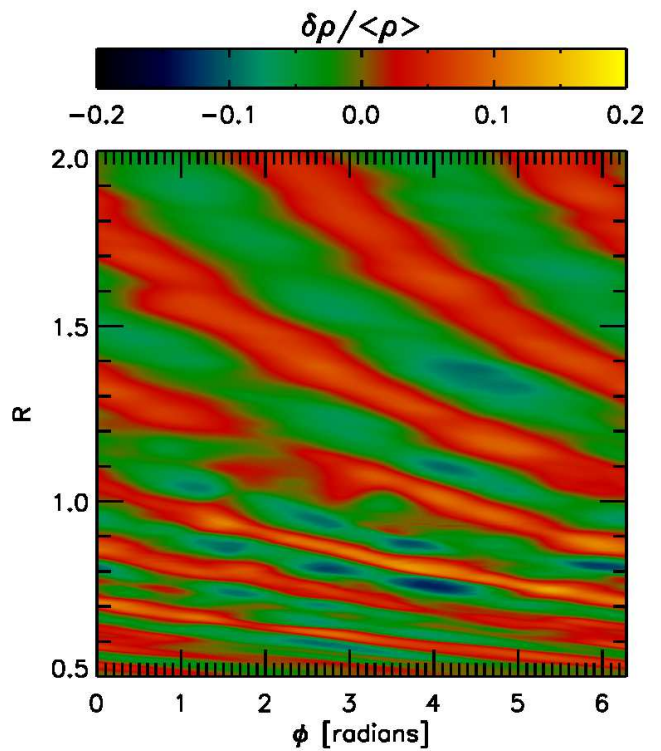}
\caption{Distribution of the normalized perturbed density $\delta\rho / \langle \rho \rangle$ in the $Z=0$ plane at $t=250$ from the CYL-F model.}
\label{fig:rhophi_2d_cyl}
\end{figure}

\begin{figure*}
\centering
\epsscale{1.15}
\plotone{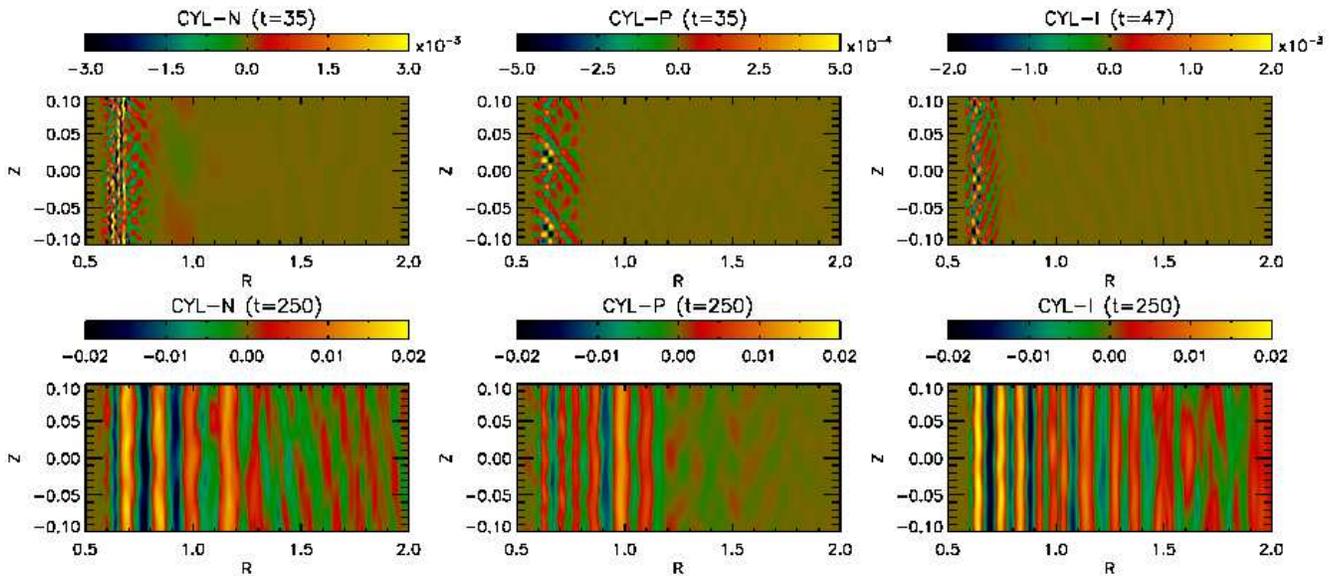}
\caption{Distributions of the vertical velocity in a vertical plane obtained with (left) NIRVANA, (middle) PLUTO, and (right) INABA3D. The upper panels present the distribution at $t=35$ to compare the early evolution of the instability with CYL-F model. For the CYL-I model, we present the data taken at $t=47$ since the growth of the instability starts at a later time (see Figure \ref{fig:kinz_cyl}). The lower panels present the distribution at the end of calculations ($t=250$).}
\label{fig:cylz-comp}
\end{figure*}

The growth rate measured during the exponential growth phase shown in Figure~\ref{fig:kinz_cyl} is measured to be $1.26~T_{\rm orb}^{-1}$. While this represents the growth associated with the superposition of numerous growing inertial modes between the radii $R=0.6 - 1$, it is likely to be dominated by the longer wavelength modes.  
The instability saturates at about $t = 50$. As shown in Figure \ref{fig:cylz-f} and \ref{fig:cylr-f}, the maximum perturbed vertical and radial velocities of the unstable inertial modes at saturation are on the order of the radial velocity of the background wave. 
This is presumably an indication of the fact that the linear instability saturates when the perturbed velocity grows large enough to disrupt the structure of the background wave, as also argued by \cite{fromang07}. 
During the non-linear phase of the instability, the checkerboard shaped cells merge together, creating alternating vertical flows. At $t=250$, the vertical flows are locally as fast as $\sim50\%$ of the sound speed. 

In Figure \ref{fig:rhophi_2d_cyl}, we present the perturbed density distribution $\delta \rho / \langle \rho \rangle$ in a horizontal ($R$, $\phi$) plane at $t=250$, where the angled brackets denote an azimuthal average and $\delta \rho = \rho - \langle \rho \rangle$.
As expected from the snapshots in the vertical plane, the spiral arms are highly perturbed, showing fragmentary structures. Moving along the azimuth at any given radius shows a clear asymmetry. 
We will further discuss the appearance of the disrupted spiral structures in Section \ref{sec:asymetric_arms}, along with their observational implications.

\subsection{Code Comparison}

In order to demonstrate the robustness of the instability we have run the same calculation with the four independent codes introduced in Section \ref{sec:codes}.
As shown in Figure \ref{fig:kinz_cyl}, clear exponential growth of the instability is observed with all codes. The growth time measured during the linear phase ranges between $\sim1.2-1.4~T_{\rm orb}^{-1}$. 
The saturated perturbed energy levels also show good agreement with each other, being within a factor $\sim3$ at $t=250$.

We note, however, that the early evolution of the perturbed kinetic energy ($t \lesssim 30$) seems to be more code dependent than the later evolution. In particular, the exponential growth seen at $5 \lesssim t \lesssim 15$ in the CYL-F model is not observed with the other codes.
As discussed earlier, this is when the small scale modes, with wavelengths comparable to grid cell size, grow in the models. 
The growth rates of such small wavelength modes are inevitably dependent on the dissipative properties of each code.

In Figure \ref{fig:cylz-comp}, we present contour plots of the vertical velocity obtained with NIRVANA, PLUTO and INABA3D.
Although the detailed evolution of the instability may slightly differ, it is clear that the same instability is being captured by all codes -- the instability exhibits a checkerboard pattern during the linear phase, and alternating vertical flows when saturated.

As mentioned in Section \ref{sec:model}, we initially add small, random, vertical velocities to seed the instability.
Adding noise puts energy into the inertial modes in the disk. When the initial random perturbation is not added, we note that the cylindrical model does not trigger the SWI with the finite volume codes PLUTO and INABA3D. 
These codes do an excellent job of maintaining the symmetry associated with the initial conditions because the conservative properties of these codes are enforced by the pairwise exchange of fluid properties between neighboring grid cells.
In the finite difference codes, however, we find that the SWI still develops when starting without the initial perturbations, presumably because machine precision level errors are introduced by the use of finite differences.

\begin{figure}
\centering
\epsscale{1.2}
\plotone{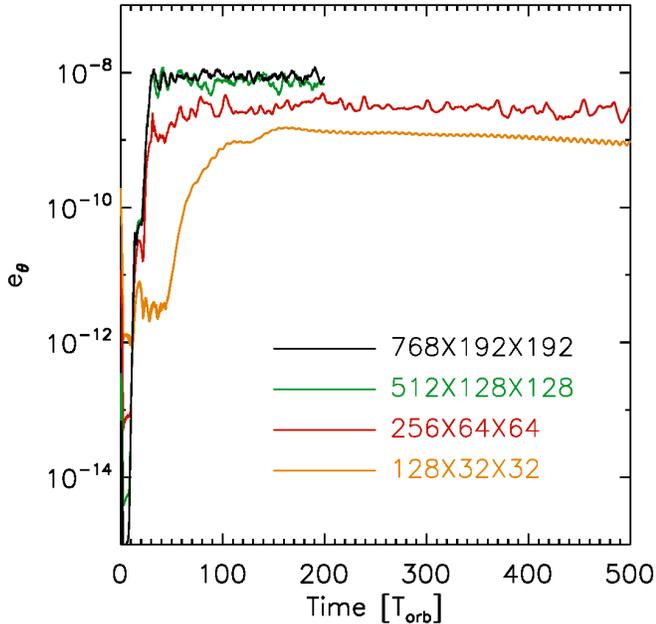}
\caption{Time evolution of the integrated meridional kinetic energy $e_\theta$. Results with four different resolutions are plotted. The plot indicates that the numerical results more or less converge at $(N_r \times N_\phi \times N_\theta) = (512\times128\times128)$ and beyond. We run the two low resolution models for a longer time to see if the saturation level eventually converges toward the value with higher resolutions, but find that the perturbed energy level does not further increase after $200~T_{\rm orb}$ in both $(N_r \times N_\phi \times N_\theta) = (256\times64\times64)$ and $(128\times32\times32)$ run.}
\label{fig:kinz_piso-ref}
\end{figure}

\begin{figure*}
\centering
\epsscale{1.11}
\plotone{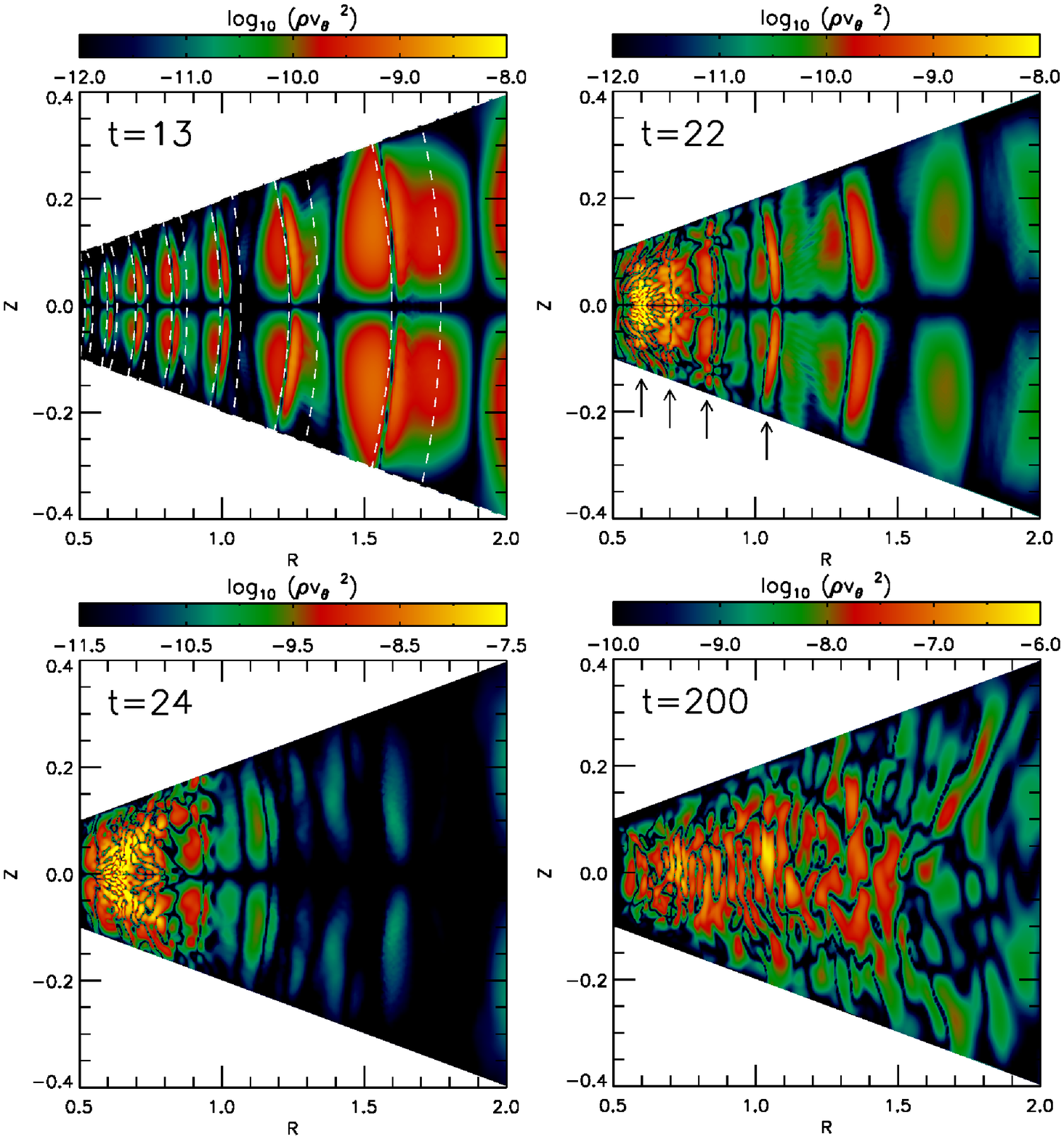}
\caption{Contour plots presenting two-dimensional distributions of the meridional kinetic energy density $\rho v_\theta^2$ in a vertical plane ($\phi=0$). The snapshots are taken (upper-left) before the spiral wave instability is triggered, (upper-right) at the beginning of the linear growth phase, (lower-left) at the middle of the linear growth phase, and (lower-right) at the end of the calculation when the instability is fully saturated. The white dashed contours in the upper-left panel show the $v_r=0$ surface. Note that the wave fronts are curved towards the midplane due to nonlinear advection of the wave. The arrows in the upper-right panel indicate the four locations where the spiral arms intersect the $\phi=0$ plane at that time.}
\label{fig:piso-r512-vtheta}
\end{figure*}

\section{VERTICALLY STRATIFIED, ISOTHERMAL DISK MODELS}
\label{sec:iso}

In this section, we present results of vertically stratified models with an isothermal equation of state.
The calculations are initiated with the vertically stratified density structure and the corresponding azimuthal velocity that satisfy the hydrostatic equilibrium, as introduced in Equations (\ref{eqn:init_den}) and (\ref{eqn:init_vel}).
We first introduce our reference model, for which the spiral wave amplitude  $\mathcal{A}=5.0\times10^{-4}$ and which has zero kinematic viscosity. The numerical resolution is $(N_r \times N_\phi \times N_\theta) = (512\times128\times128)$.
The effects of varying the numerical resolution, spiral potential amplitude, and viscosity will be discussed in the following sections. 

\begin{figure*}
\centering
\epsscale{1.1}
\plotone{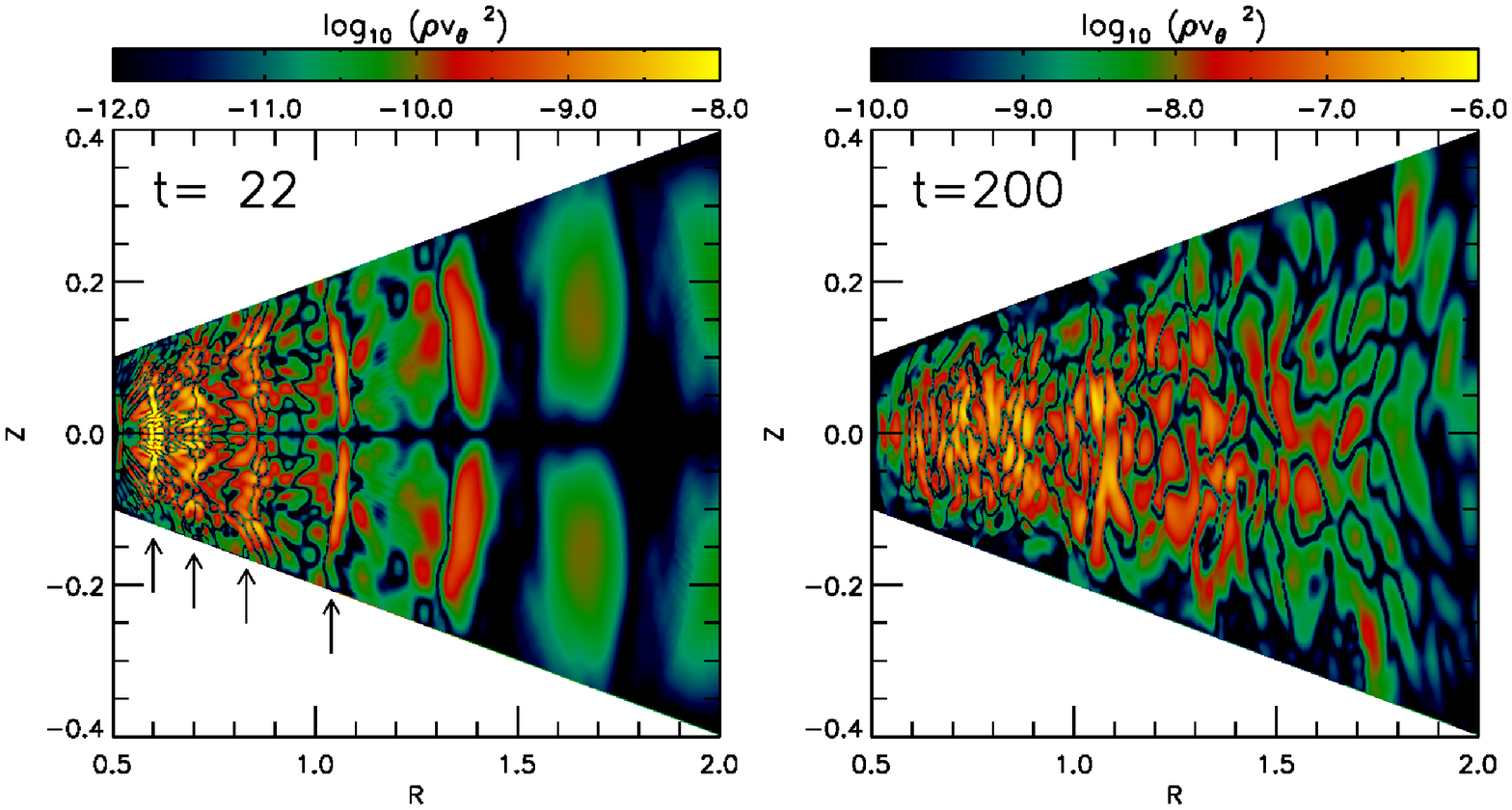}
\caption{Contour plots of the meridional kinetic energy density $\rho v_\theta^2$ in a vertical plane ($\phi=0$), obtained with $(N_r \times N_\phi \times N_\theta) = (768\times192\times192)$ grid cells.}
\label{fig:piso-r768}
\end{figure*}

As discussed in Section \ref{sec:background}, we expect that the spectrum of inertial modes that can be excited by the spiral waves will be dense, and therefore it is difficult to identify individual growing modes during the simulations. 
The best way of doing this would be to compute the eigenfunctions associated with these modes, such that they could be seeded in the initial conditions and their growth rates measured. 
We take a different approach by using white noise to perturb the initial conditions, thereby adding energy to the full spectrum of inertial modes, and ensuring the robustness of the SWI by examining the triggering of the instability under a variety of disk conditions, instead of identifying and characterizing individual unstable modes.

\subsection{A Reference Run (R512)}
\label{sec:ref}

In Figure \ref{fig:kinz_piso-ref}, we present the time evolution of the integrated meridional kinetic energy $e_\theta$.
Contour plots of the meridional kinetic energy density $\rho v_\theta^2$ are plotted in Figure \ref{fig:piso-r512-vtheta}.
We make use of the meridional kinetic energy density $\rho v_\theta^2$ for the stratified models, instead of the meridional velocity, in order to better illustrate the development of the instability in the main body of the disk.

As shown in Figure \ref{fig:kinz_piso-ref}, $e_\theta$ increases rapidly very early in the simulations at $10 \lesssim t \lesssim 15$.
This initial increase does not represent the growth of unstable inertial modes.
As shown in the first panel of Figure \ref{fig:piso-r512-vtheta}, the spiral waves are not perpendicular to the disk midplane in the stratified models, but instead have a concave shape that develops as they propagate towards the star. 
As discussed in Section~\ref{sec:background}, this seems to arise as a nonlinear effect due faster advection of the wave at higher altitudes in the disk because the initial planar wave fronts move through steeply decreasing density profiles there, whereas the density increases near the midplane as the wave moves inwards. 
The curvature results in a loss of vertical hydrostatic equilibrium in the vertical direction, giving rise to the meridional kinetic energy.

The exponential growth associated with the SWI starts at $t \sim 20$, and $e_\theta$ saturates at $t \sim 30$.
The second and third panel of Figure \ref{fig:piso-r512-vtheta} show the linear growth and saturation of unstable modes at the innermost spiral arm. 
The unstable modes grow faster at smaller radii as shown, because of the higher mode frequencies and growth rates closer to the central object.
When the instability is saturated ($t=200$), the disk ends up in a complicated quasi-steady turbulent state.
We discuss the turbulence produced via the SWI and its implications for angular momentum transport and vertical mixing later in Section \ref{sec:transport}.

\subsection{Effect of Numerical Resolution}

In order to test the effect of numerical resolution, we run the reference model with four different resolutions: $(N_r \times N_\phi \times N_\theta) = (128\times32\times32), (256\times64\times64), (512\times128\times128)$, and $(768\times192\times192)$.
Increasing the numerical resolution should allow the growth of smaller scale inertial modes. 

As seen in Figure \ref{fig:kinz_piso-ref}, the growth rate and the saturated perturbed energy are reasonably well converged at $(N_r \times N_\phi \times N_\theta) = (512\times128\times128)$ and beyond. 
As expected, the highest resolution run displays the highest level of perturbed energy in the saturated state on average, because of the excitation of smaller scale modes, but the difference between the $(512\times128\times128)$ and $(768\times192\times192)$ runs is modest, indicating that the modes containing the largest amount of energy are captured in the lower resolution calculation. 
We note that there is a significant difference between the two higher resolution cases and the $(256\times64\times64)$ and $(128\times32\times32)$ cases, with the lowest resolution run in particular showing a much lower growth rate and saturation level.  

In Figure \ref{fig:piso-r768}, we present contour plots of the meridional kinetic energy density for  R768 model at $t=22$ and 200. 
Compared to the results obtained with $(512\times128\times128)$ grid cells presented in Figure \ref{fig:piso-r512-vtheta}, one can see more fine scale structures developing at the beginning of the instability ($t=22$).
It is also apparent that, with the higher resolution, the unstable modes have grown faster at the third and fourth innermost spiral arms.
In the R768 model, we find that the spiral waves create a larger density enhancement at the wave fronts than in the reference model: at $r=0.6$, the spiral waves induce $39\%$ of density enhancement in the midplane in R768 model, whereas the enhancement is $37\%$ in the reference model.
Since the growth rate of unstable modes is an increasing function of the wave amplitude \citep[][and see below]{fromang07}, the stronger perturbations lead to faster growth of the unstable modes, in addition to there being more modes excited at high resolution.

\begin{figure}
\centering
\epsscale{1.2}
\plotone{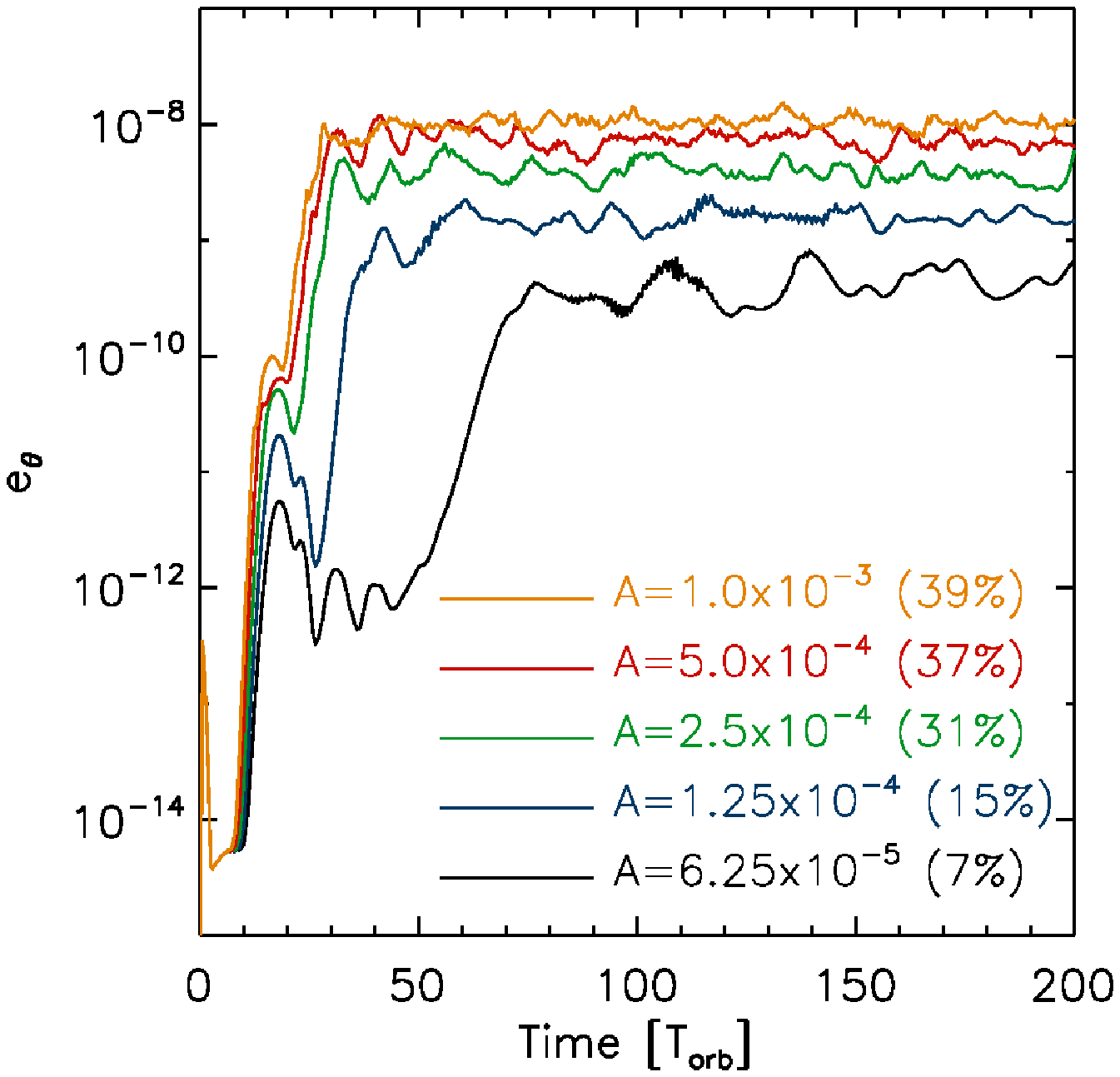}
\caption{Time evolution of the meridional kinetic energy $e_\theta$ for different potential amplitude $\mathcal{A}$. The percentage in the parentheses indicate the density enhancement induced by the imposed spiral waves at $r=0.6$ (refer to the right panel of Figure \ref{fig:waves}).}
\label{fig:kinz_piso_amp}
\end{figure}

\begin{figure}
\centering
\epsscale{1.2}
\plotone{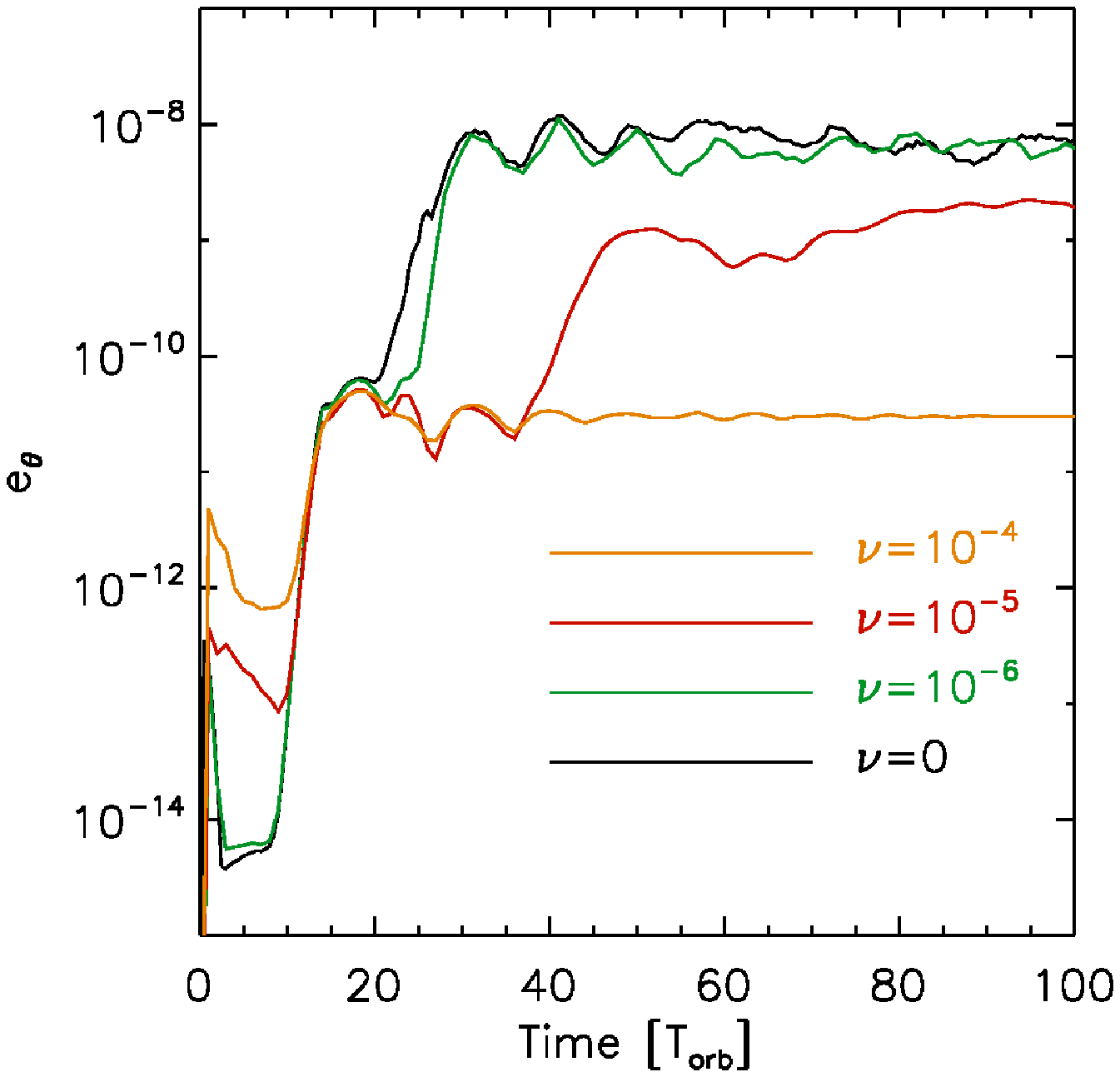}
\caption{Time evolution of the meridional kinetic energy $e_\theta$ for different values of the kinematic viscosity. Note that the spiral wave instability operates with $\nu \leq 10^{-5}$, while the instability is suppressed when $\nu=10^{-4}$.}
\label{fig:kinz_piso_visc}
\end{figure}

\subsection{Effect of Spiral Potential Amplitude}
\label{sec:pstrength}

In this section, we explore the effect of varying the imposed spiral potential strength. 
We increase the spiral potential amplitude by factors of two from the smallest value, $\mathcal{A}=6.25\times10^{-5}$, up to the largest value $\mathcal{A}=1.0\times10^{-3}$.

Figure \ref{fig:kinz_piso_amp} shows the time evolution of $e_\theta$. 
It is immediately obvious from the figure that the instability has a higher perturbed energy level with stronger spiral waves.
During the initial phase when the spiral waves propagate into the inner disk ($t \lesssim 20$), stronger spiral waves create more vertical motion.
In addition, the linear growth phase begins at an earlier time with a stronger spiral amplitude.
In the case with $\mathcal{A}=6.25\times10^{-5}$, for example, the instability starts at about $t=50$, while in the reference model the instability starts almost immediately after the initial propagation of the waves.

From $\mathcal{A} = 6.25 \times 10^{-5}$ to $\mathcal{A} = 2.5 \times 10^{-4}$, the density enhancement doubles when the spiral potential amplitude is doubled.
However, the density enhancement induced by the spiral waves does not linearly increase with the potential strength when $\mathcal{A} > 2.5 \times 10^{-4}$.
Comparing $\mathcal{A} = 2.5 \times 10^{-4}$ model with $\mathcal{A} = 10^{-3}$ model, the density enhancement increases only by about $25\%$, presumably as a result of nonlinear dissipation. 
The nonlinear dissipation is observed over a broad disk region ($r \lesssim 1.6$), and interestingly, it is more significant at smaller radius.

\begin{figure*}
\centering
\epsscale{1.1}
\plotone{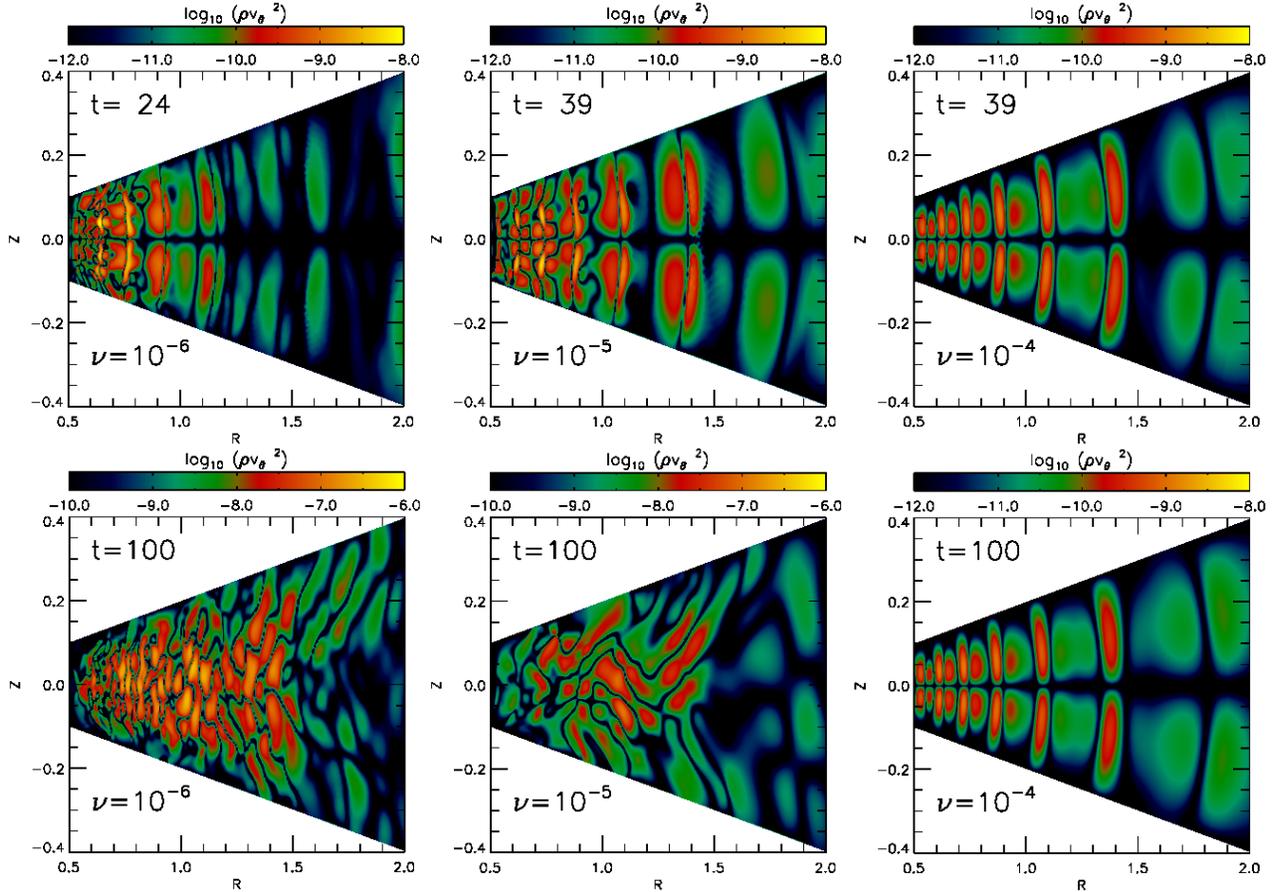}
\caption{Contour plots showing two-dimensional distributions of the meridional kinetic energy density $\rho v_\theta^2$ in the vertical plane with different kinematic viscosities. The upper panels show the distributions at the beginning of the linear growth phase, and the lower panels show the distributions at $t=100$. Note that the instability operates with (left) $\nu=10^{-6}$ and (middle) $\nu = 10^{-5}$, but is suppressed with (right) $\nu = 10^{-4}$.}
\label{fig:piso_visc}
\end{figure*}

\subsection{Effect of Viscosity}

We add a constant kinematic viscosity to our reference model to study the triggering and the evolution of the SWI in the presence of viscous dissipation.
We test with three different kinematic viscosities: $\nu=10^{-6}, 10^{-5}$, and $10^{-4}$.
In the canonical $\alpha$ prescription of \citet{shakura73}, the kinematic viscosity $\nu$ corresponds to 
\be
\alpha(R) = 1.1\times10^{-3} \left( {\nu \over 10^{-6}} \right) \left( {R \over 0.5} \right)^{-1.5}.
\en
If we assume that radii are measured in units of AU, then a value of $\nu=10^{-5}$ corresponds to $\alpha=3.9 \times 10^{-3}$ at 1 AU, and $\alpha=3.5 \times 10^{-4}$ at 5 AU.
In Figure \ref{fig:kinz_piso_visc}, we plot the time evolution of the integrated meridional kinetic energy.
As the plot indicates, the SWI operates with $\nu \leq 10^{-5}$, but is completely suppressed when $\nu = 10^{-4}$.
The linear growth of the instability begins at a later time with a larger viscosity, probably because the viscosity damps the smaller scale fastest growing modes. 
Also, the saturated energy level is reduced with a larger viscosity, for the same reason, although the difference between $\nu=0$ and $\nu=10^{-6}$ models is only marginal, suggesting that numerical diffusion operates with a value that is moderately below $\nu=10^{-6}$ in the inviscid model.

\begin{figure}
\centering
\epsscale{1.2}
\plotone{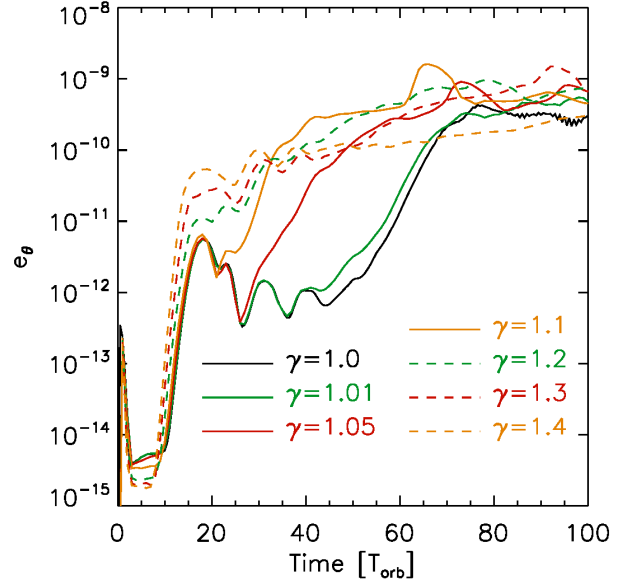}
\caption{Time evolution of the meridional kinetic energy $e_\theta$ for different adiabatic indices $\gamma$.}
\label{fig:kinz_piso_gam}
\end{figure}

\begin{figure*}
\centering
\epsscale{1.18}
\plotone{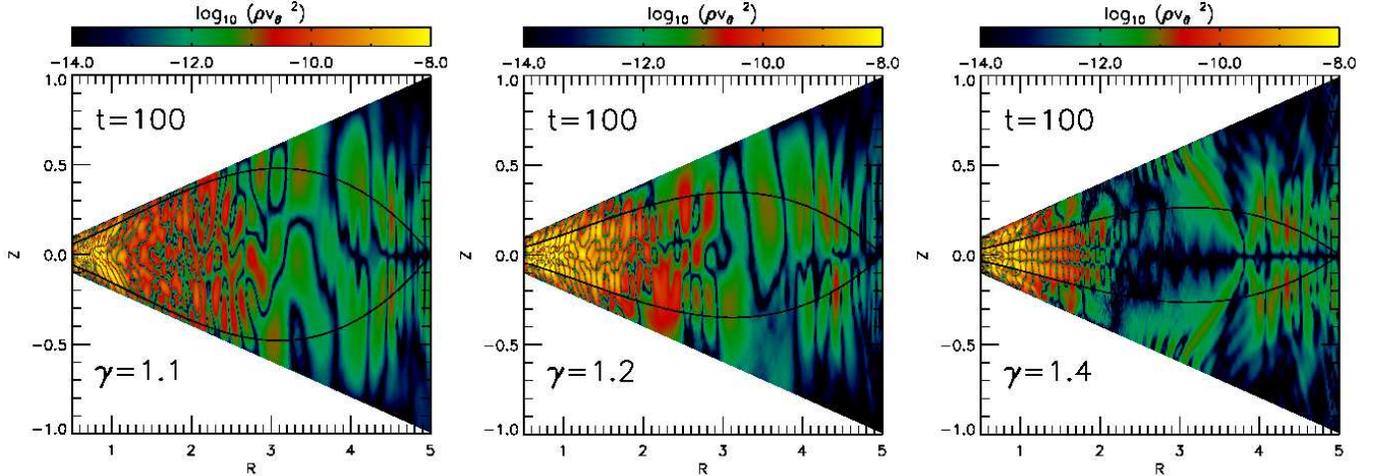}
\caption{Contour plots of the meridional kinetic energy density $\rho v_\theta^2$ for adiabatic models with different $\gamma$ values. The black curves indicate where the local buoyancy frequency equals to a half of the local Doppler-shifted frequency of the spiral waves: $N^2 = ({\omega}_{\rm s}/2)^2$. The results from the $\gamma=1.4$ run demonstrate the fact that the region where the SWI develops is confined to the regions where $N^2 \lesssim (\omega_s/2)^2$.}
\label{fig:adia}
\end{figure*}

Contour plots of meridional kinetic energy density are shown in Figure \ref{fig:piso_visc}.
As inferred from the energy evolution, one can see that the SWI operates with $\nu \leq 10^{-5}$.
With $\nu=10^{-4}$, we do not find any signatures of the instability growing throughout the duration of the calculation.
Comparing with the results of the reference model at $t=24$ presented in Figure \ref{fig:piso-r512-vtheta}, we find that the small scale unstable modes are absent with $\nu=10^{-6}$.
With $\nu=10^{-5}$, small scale modes are further suppressed, and the unstable modes that have grown in the model have noticeably larger length scales than in the $\nu=0$ and $\nu=10^{-6}$ models. 
\cite{fromang07} show that, except for the largest wavelength inertial modes, the growth rates are almost constant. Given that we expect modes to be viscously damped when the damping rate exceeds the growth rate, and the viscous damping rate scales as  $\sim k^2 \nu$, this demonstrates that we expect the high $k$ (small $\lambda$) modes to be preferentially damped.

We emphasize that the instability operates with fairly large viscosity of $\nu = 10^{-5}$. 
In terms of the canonical $\alpha$ description, $\nu = 10^{-5}$ corresponds to $\alpha \sim 0.008$ at $R=0.6$ (right outside of the inner damping zone) and $\alpha \sim 0.001$ at $R=2$.
The fact that the spiral wave instability operates in quite dissipative disks may be important for disks that sustain fully developed MHD turbulence, such as the accretion disks around cataclysmic variables and/or black holes. 
Furthermore, the growth rate of the instability depends on the amplitude of the spiral wave, so the value of $\nu$ for which the SWI operates will depend on the incoming wave amplitude.

\section{VERTICALLY STRATIFIED, ADIABATIC DISK MODELS}
\label{sec:adia}

In this section, we adopt an adiabatic equation of state to investigate the influence of the thermodynamics on the growth and development of the SWI.
We consider six different adiabatic index values: $\gamma=1.01, 1.05, 1.1, 1.2, 1.3, 1.4$.
By using different values of $\gamma$, we mimic a disk that experiences cooling at different rates. Reducing $\gamma$ acts as if the disk has more efficient cooling.
Using the broad range of $\gamma$ values, we aim to demonstrate that the SWI operates in both optically thin regions (isothermal) and optically thick regions (adiabatic).
As we shall show below, the disks become more susceptible to the SWI with stronger adiabatic responses (i.e. larger $\gamma$ value).
In contrast to the reference model that we have presented earlier in this paper, we use a smaller spiral potential amplitude of $\mathcal{A} = 6.25\times10^{-5}$ to allow us to capture the growth of the instability at early times.

In Figure \ref{fig:kinz_piso_gam}, we present the time evolution of the integrated meridional kinetic energy.
When the spiral waves initially propagate into the inner disk ($t \lesssim 20$), a more rapid and higher amplitude growth of $e_\theta$ is observed with a larger $\gamma$ value, and it is clear that the SWI triggers at earlier times with larger $\gamma$ values.
For example, the linear growth of the instability begins at $t \sim 45$ with $\gamma=1.01$, but at $t \sim 30$ with $\gamma=1.05$ and at $t \sim 25$ with $\gamma=1.1$. 
For $\gamma=1.4$, it is difficult to disentangle the initial growth of the perturbed kinetic energy due to the spiral wave propagation from the growth of the instability.

As we discussed in Section~\ref{sec:background}, the Brunt-V\"ais\"al\"a frequency varies over a wide range of values over height in adiabatic disks, with the range of values covered being a function of the value of the effective $\gamma$ adopted. 
This suggests that at each radius in an adiabatic disk, the spiral modes will have a larger number of inertial modes with which to resonantly interact, with the number being larger for larger values of $\gamma$. 
We might therefore expect to see a more rapid growth of the SWI for the larger values of $\gamma$, and this expectation is borne out by the simulation results.

We also pointed out in Section~\ref{sec:background} that there can be forbidden regions for the SWI in adiabatic disks.
The existence of these can be inferred from Equation (\ref{eqn:kZi2}), which shows that the vertical wave number has no physical solution in disk regions where  $({\omega}_{\rm s}/2)^2 \lesssim N^2$ in the large $n$ limit. 
Although in global models, such as those we present here, the analysis of the properties of the inertial waves should also be global and not local, the analysis of the vertical structure of inertial mode eigenfunctions by \cite{lubow93} also shows that the energy in the modes is confined near the midplane, and the modes do not propagate in regions where the inertial mode frequencies $\omega_{\rm i}^2 < N^2$. 

In Figure \ref{fig:adia}, we display contour plots of the meridional kinetic energy density at the end of calculations ($t=100$) with $\gamma=1.1$, 1.2, and 1.4. 
In the figure, the black curves connect the disk regions where the local Brunt-V\"ais\"al\"a frequency matches to a half of the Doppler-shifted spiral wave frequency: $N^2 = ({\omega}_{\rm s}/2)^2$.
As discussed above, the region surrounded by the black curves is where the inertial  waves are expected to be confined.
The forbidden region is not clearly seen in the $\gamma = 1.1$ and 1.2 models (although there is a strong hint of it in the latter model), because the $N^2 = ({\omega}_{\rm s}/2)^2$ region is located near the meridional boundary, and acoustic waves excited by the SWI are likely to be present there along with vertical motions induced by the spiral waves. 
We note, however, the strong signature of inertial mode confinement near the midplane in the $\gamma=1.4$ model. 

Finally, we note that varying the adiabatic index affects the shape and strength of spiral waves.
As $\gamma$ is increased, spiral waves are more openly wound, due to the larger sound speed, and the density enhancement at wave fronts is reduced. 
These, in turn, can have influence on the growth rate of the SWI and the saturated energy level, although the existence of the SWI in adiabatic disks is clearly robust. 
At $r=0.6$, for example, the density perturbation $\rho / \langle \rho \rangle$ before the SWI sets in is $\sim 7~\%$ with $\gamma=1.0$, $\sim 4~\%$ with $\gamma=1.1$, and only $\sim 1~\%$ with $\gamma=1.4$.
The reduction in density enhancement at spiral wave fronts for larger values of $\gamma$ may explain why the saturated state of the model with $\gamma=1.4$ is somewhat smaller than in the other models (as seen in Figure~\ref{fig:kinz_piso_gam}), in spite of the obvious faster growth rate associated with this run.

\begin{figure}
\centering
\epsscale{1.15}
\plotone{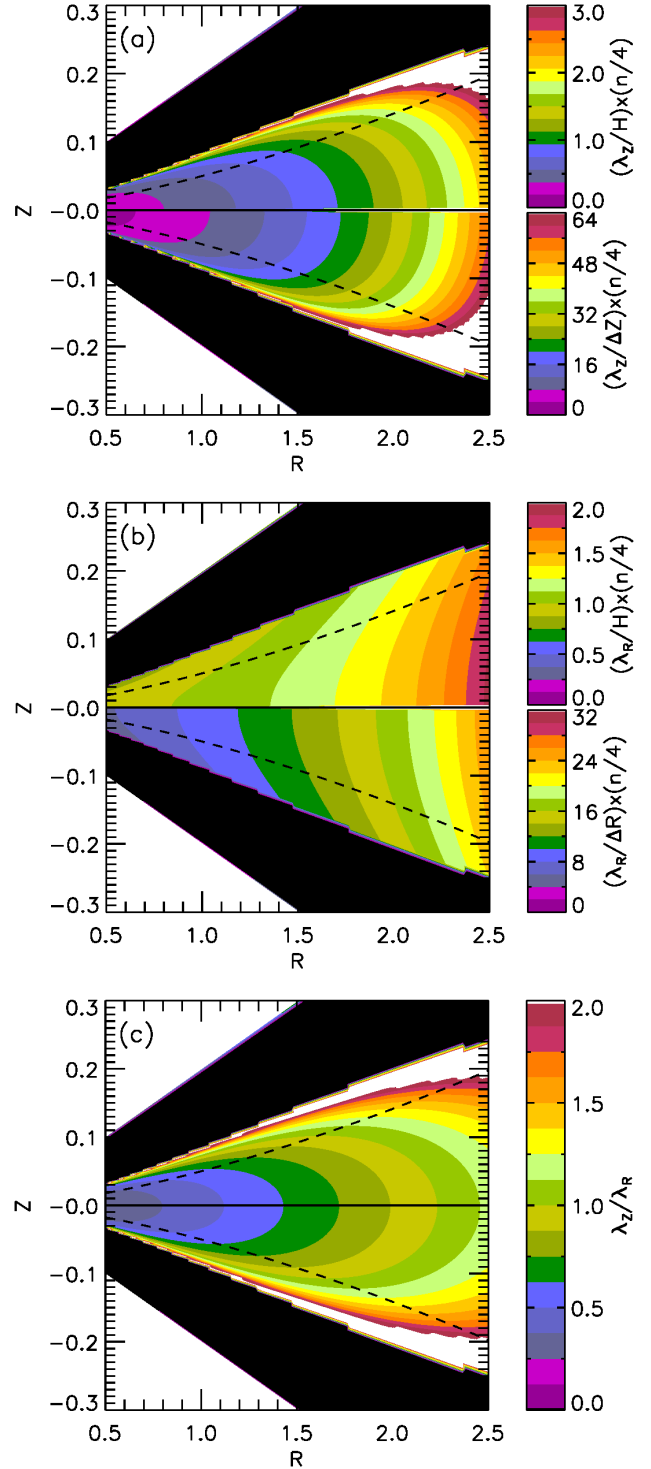}
\caption{(a) Contour plots of the vertical wavelength of the inertial modes $\lambda_Z$ that are subject to the resonant interaction with the imposed spiral waves assumed in this work. (b) Same as (a) but for the radial wavelength $\lambda_R$. (c) The ratio of vertical to radial wavelength of the inertial modes. In all panels, the dashed curves indicate where $Z=\pm 1 H$. The black regions are where the dispersion relation does not have a physical solution. In (a) and (b), the upper half of the disk shows the wavelength in units of disk scale height at each radius, whereas the lower half of the disk shows the wavelength in units of vertical and radial grid cell size at each position. We use the $n=4$ mode as a representative, but the numbers on the colorbars can be linearly scaled to other modes as noted in the labels.}
\label{fig:dr_sph}
\end{figure}

\section{DISCUSSION}
\label{sec:discussion}

Our results suggest that the SWI can operate under a broad range of disk conditions.
In addition, the instability presumably operates regardless of the origin of spiral waves, as long as the waves maintain a fixed period until the instability develops\footnote{We have confirmed that the SWI develops for the spiral waves driven by a companion in a disk.}.
This implies a potential significance of the SWI in various astrophysical disks.

\subsection{Why Has the Instability Not Been Reported Previously?}

At this point, one might wonder why the SWI has not been reported previously if it operates under a broad range of disk conditions as we claim.

First of all, it is possible that the SWI has been present in previous studies, but masked by turbulence driven via other processes.
For instance, the recent study of nonlinear spiral waves in a disk excited by a massive planet by \cite{lyra16} used high resolution in an essentially inviscid disk, but the violence of the disk response to the strong nonlinear forcing, combined with the presence of convective motions, probably masked the presence of the SWI in that study. 
Also, we speculate the SWI might have been present in the recent 3D MHD calculation of accretion disks in cataclysmic variable (CV) systems by \citet{ju16}.
In this study, however, the SWI must be obscured by the vigorous MRI turbulence.

Considering the reference run R512, this was computed using a logarithmic radial grid, where the grid cell size at $r=0.5$ was $\Delta r = 0.002934$. Similarly, the run undertaken by NIRVANA, described in the appendix, used a uniform radial mesh with the same grid spacing, corresponding to $N_r =3200$ grid cells. 
This is higher resolution than is usually undertaken in global simulations of giant planets embedded in protoplanetary disks, for example, although high resolution in the vicinity of the planet is often achieved with local mesh refinement \citep[e.g][]{kley01,klahr06,gressel13,szulagyi14}.  
Viscosity, with $\nu \ge 10^{-5}$, is also often included in these models, which has the effect of partially damping the instability. 
We note that simulations with low-mass planets are unlikely to show growth of the SWI within the typical durations of three-dimensional calculations, since the smaller the mass of companion, the weaker the instability and the longer it takes for it to develop (see Figure \ref{fig:kinz_piso_amp}).
In addition, the primary and secondary arm separation is known to be less than $180^\circ$ for low-mass planets \citep[e.g.][]{fung15,zhu15}, in which case the resonance between the spiral arms and inertial modes is not exact.
In this case, the SWI may grow at a reduced rate, requiring a longer simulation duration to observe the instability develop.

In Figure \ref{fig:dr_sph}, we present two-dimensional maps showing the vertical and radial wavelengths of the inertial modes that satisfy the resonance conditions with the imposed spiral waves. 
The wavelengths are calculated using Equations (\ref{eqn:k-spiral}) and (\ref{eqn:kZi2}), using the background disk structure and computational domain used in the adiabatic simulation labelled as GAM4 in Table~\ref{tab:parameters}.
Purely for the purposes of illustration, we present maps for the inertial modes predicted to arise when  $n=4$, where $n$ is defined as $k_{R,{\rm i}} = n k_{R,{\rm s}}$.
However, since the second term in Equation (\ref{eqn:kZi2}) is typically smaller than the first term, the contour plots can be linearly scaled to other modes\footnote{We confirmed that this is invalid only in a very narrow region near the boundary between the SWI forbidden and permitted regions where $({\omega}_{\rm s}/2)^2 \simeq N^2$.} as indicated in the color bar labels.

The maps imply that there is a resolution limit on which modes can be represented by the computational grid. 
For $n=4$, the midplane region interior to $R \sim 0.9$ is poorly resolved by $\lesssim 8$ grid cells in both the vertical and radial directions, indicating that these inertial modes cannot be well represented on the mesh there. 
Smaller values of $n$ are hence expected to be associated with the modes seen to grow in this region, in agreement with the results described in Section~\ref{sec:cyl}. 
Moving out to the midplane region beyond $R=1$, the vertical and radial wavelengths of the excited modes start to exceed $\sim 10$ grid cell spacings, suggesting that these modes can be represented on the grid.

\subsection{Implications}

\subsubsection{Angular Momentum Transport}
\label{sec:transport}

\begin{figure}
\centering
\epsscale{1.2}
\plotone{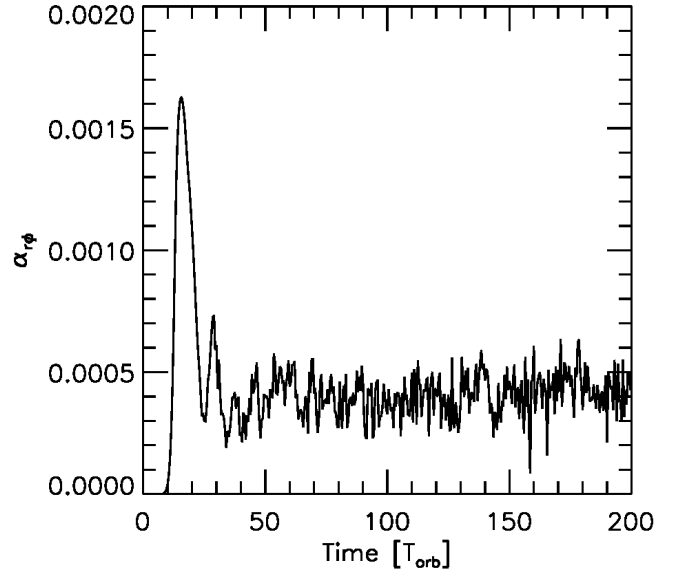}
\caption{Time evolution of Shakura-Sunyaev parameter $\alpha_{r \phi}$, averaged over $\theta$ in between $r=0.5$ and 2.}
\label{fig:alpha}
\end{figure}

In order to estimate the rate of angular momentum transport induced by the turbulent flow arising from the SWI, we calculate the Shakura-Sunyaev stress parameter $\alpha_{r \phi}$ for the reference model R512 according to $\alpha_{r\phi}(r, \theta) \equiv {{\langle \rho \delta v_r \delta v_\phi \rangle} / \overline{P}(r) }$  where $\overline{P}(r)$ is a density-weighted mean pressure at radius $r$. 
The simple arithmetic average of $\alpha_{r\phi}$ over $\theta$ in between $r=0.5$ and 2 is plotted in Figure~\ref{fig:alpha} as a function of time.
We see that the advected flux of angular momentum due to the spiral waves at the beginning of the simulation gives rise to an apparent Reynolds stress $\alpha_{r\phi} \sim 0.0016$. 
This is not the accretion stress experienced by the disk, but instead represents the negative angular momentum flux associated with the spiral wave as it propagates inwards, and only that fraction of the wave angular momentum that is deposited in the gas through wave dissipation acts to drive accretion. 
Once the SWI develops we see that the correlated velocity fluctuations generate a sustained accretion stress $\alpha_{r\phi} \sim 5 \times 10^{-4}$. 
This value is naturally smaller than that associated with the unattenuated spiral wave because the SWI operates over a range of radii in the disk, and the energy and angular momentum associated with the wave are injected into the disk matter over this radial range via the breakdown into turbulence.

\begin{figure*}
\centering
\epsscale{1.15}
\plotone{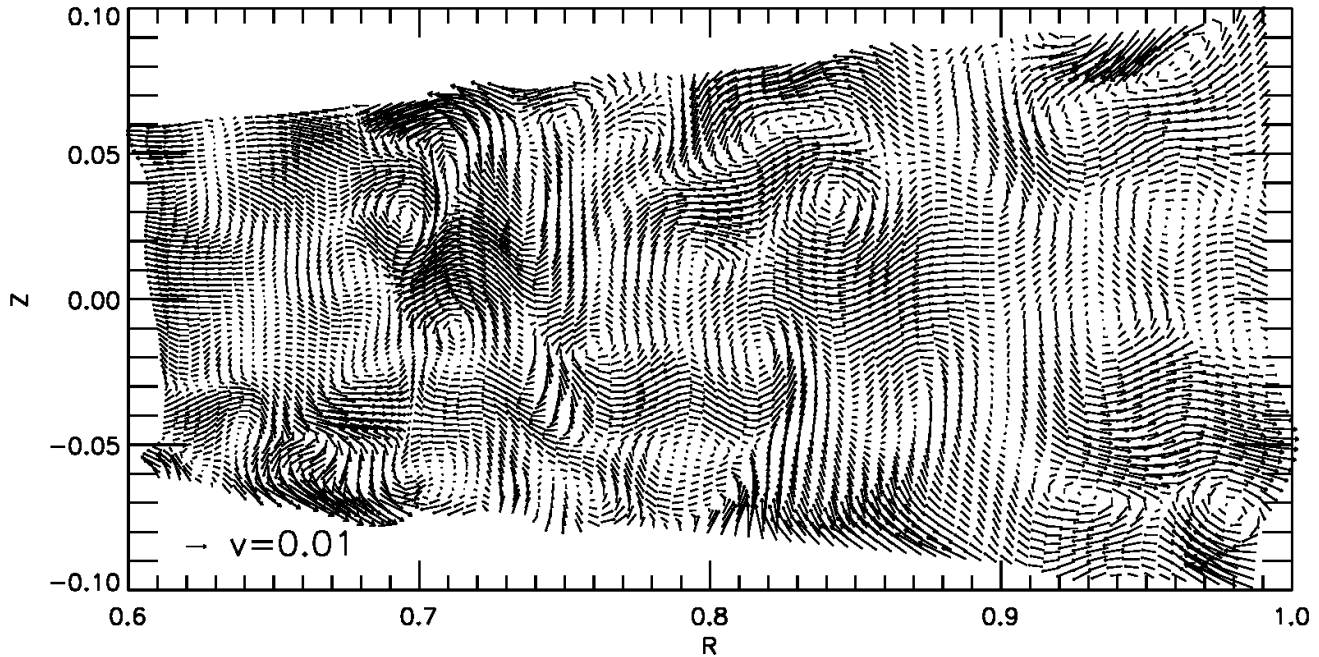}
\caption{Velocity vectors in a $R-Z$ plane at $t=200$ from the R512 model. The map shows only $Z= -2H$ to $+2H$ due to large velocity in the upper layer. Note the turbulent eddies generated by the spiral wave instability. A velocity vector with $v=0.2c_s = 0.01$ is shown on the lower-left corner.}
\label{fig:velvectors}
\end{figure*}

\subsubsection{Vertical Mixing}
\label{sec:mixing}
In Figure \ref{fig:velvectors}, we present velocity vectors in a vertical plane at the end of the reference run.
As can be seen, the SWI creates a set of turbulent eddies that will be effective at inducing vertical mixing of the gas, and any dust particles that it contains. 
It is also the case that forces arising from the turbulent density fluctuations will induce stochastic migration and eccentricity/inclination excitation of larger bodies, such as planetesimals and protoplanets, in a protoplanetary disk in which nonlinear spiral waves propagate, similar to what is observed in disks that sustain magneto-rotational turbulence \citep{nelson04,nelson05}. We estimate the vertical diffusion coefficient associated with the correlated vertical velocity fluctuations using the approximation $\mathcal{D}_Z = \langle v_Z ^2\rangle \tau_{\rm corr}$, where $\tau_{\rm corr}$ is the correlation time of the vertical velocity fluctuations, $v_Z$ \citep{fromang06}. 
In obtaining an estimate for $\tau_{\rm corr}$, we generate a time series for $v_Z$ at different locations in the disk, defined by the region $|Z| < 1H$ at $r=1$, and filter these time series to remove the sinusoidal component induced by the spiral wave. 
We then compute the autocorrelations of these time series, to which we fit the function $\exp{(-t/\tau_{\rm corr})}$, leading to an average of the measured correlation times $\tau_{\rm corr} \simeq 0.12$T$_{\rm orb}$. 
Using natural units, the diffusion coefficient is estimated to be $\mathcal{D}_Z \simeq 9 \times 10^{-6}$, such that the time scale for vertical mixing $t_{\rm mix} = H^2/\mathcal{D}_Z \simeq 44$ orbits. 

Particles in the Epstein drag regime have a settling time that can be expressed as $t_{\rm settle} = (2 \pi t_{\rm s} \Omega)^{-1}$ orbits. $t_{\rm s}$ is the stopping time defined by $t_{\rm s} = (\rho_{\rm grain}  a )/(\rho v_{\rm therm})$ \citep{weidenschilling77}, where $\rho_{\rm grain}$ is the internal density of the grain particles (typically $\sim 2$~g cm$^{-3}$), $a$ is the grain size and $v_{\rm therm}$ is the thermal velocity of gas molecules. 
Assuming that particles are significantly mixed by the turbulence when $t_{\rm mix} \sim t_{\rm settle}$, we expect this to occur for particles with Stokes numbers ${\rm St} \equiv t_{s} \Omega \le 0.004$. 
This corresponds to a particle size $a \sim 2$ cm at 1 AU and $a \sim 2$ mm at 5 AU in the midplane of a minimum mass solar nebula \citep{hayashi81}.\footnote{We note that the mean free path of molecules is $\sim 1$~cm at 1 AU, so using the Epstein drag formula here is only marginally justified for 2~cm-sized pebbles.}

\subsubsection{Non-axisymmetric Spiral Features}
\label{sec:asymetric_arms}

Recent near-IR scattered light observations have revealed complex spiral arm morphologies in some protoplanetary disks \citep[e.g.][]{garufi13,benisty15,garufi16}, thanks to the advent of extreme adaptive optics.
The scattered light is believed to trace (sub-)$\mu$m-sized dust grains in the disk atmosphere since protoplanetary disks are highly optically thick at near-IR wavelength. 
Interestingly, even in nearly face-on systems exhibiting well-defined spiral arms, the brightness of the spiral arms at the same distance from the central object significantly differ from each other (e.g. SAO~206462; \citealt{garufi13}, MWC~758; \citealt{benisty15}).

\begin{figure*}
\centering
\epsscale{1.1}
\plotone{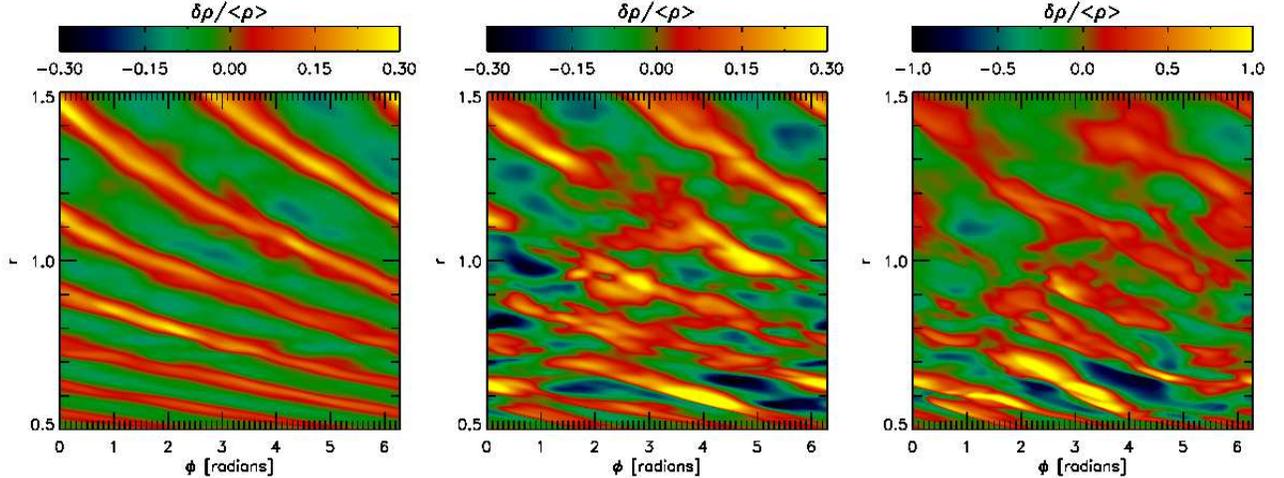}
\caption{Distribution of $\delta\rho / \langle \rho \rangle$ at $t=200$ in the (left) $Z=0$ plane, (middle) $Z=2H$ plane, and (right) $Z=3H$ plane for R512 model.}
\label{fig:rhophi_2d}
\end{figure*}

\begin{figure}
\centering
\epsscale{0.99}
\plotone{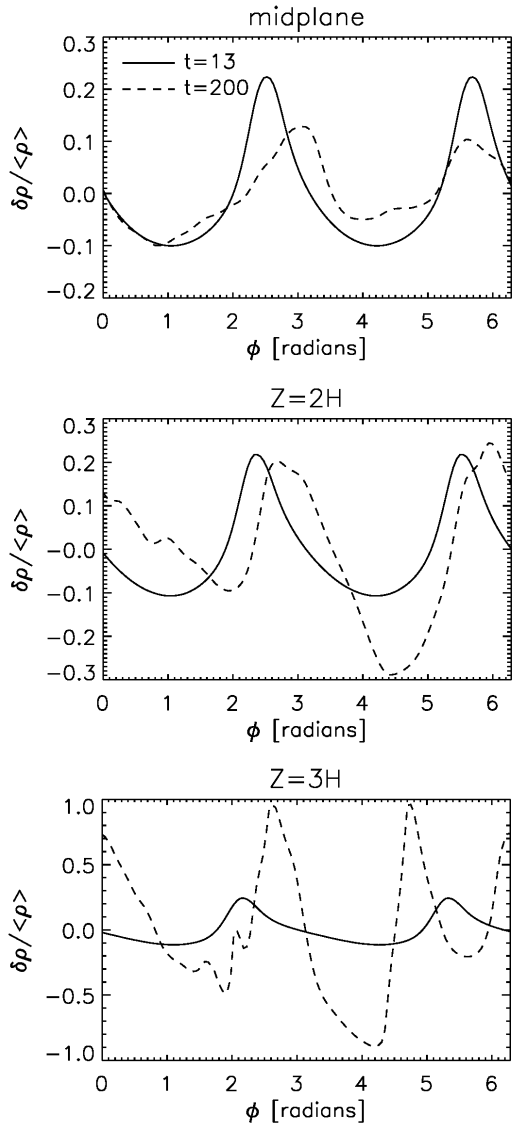}
\caption{Azimuthal distributions of the perturbed density $\delta\rho / \langle \rho \rangle$ at different heights in the disk at $r=0.65$. The solid curves show the distributions before the SWI sets in, and the dashed curves show the distributions when the instability is saturated.}
\label{fig:rhophi}
\end{figure}

In Figure \ref{fig:rhophi_2d}, we present two-dimensional distributions of the azimuthal density variation at various heights in the disk, and in Figure~\ref{fig:rhophi}, we show how the perturbed density varies with azimuth at different disk heights.
We see that spiral waves are disturbed by the instability at all heights, but with more prominent disruption occurring higher in the disk. 
This will obviously have significant implications when attempting to predict the appearance of spiral waves for comparison with observations \citep[e.g.][]{dong16}

\subsection{Future Work}

Our results suggest that the SWI is robust, and possibly operates under a variety of disk conditions.
The simplified disk models assumed in the present work help to illuminate the nature of the instability, but the significance of the instability will have to be tested with more realistic disk models.

The influence of thermodynamic effects on the instability will need to be tested using a more realistic model for the thermal evolution, including processes such as radiation transport and external irradiation. 
We have shown that the SWI is likely to operate in both optically thin (isothermal) and thick (adiabatic) disk regions. 
On the other hand, the outcome of the instability is quite sensitive to the disk thermal model, as we showed by varying the adiabatic index. 
Real disks are likely to behave adiabatically near the midplane and isothermally near the surface, so the outcome of t-he instability in such disks will be more complicated than shown by the models we have presented here. 
In particular, the forbidden zones, where inertial waves cannot be excited, will depend on the detailed thermodynamics due to the dependency on the local Brunt-V\"as\"al\"a frequency. 

Our simulations have shown that the SWI can operate in viscous disks, suggesting that it might also operate in disks that sustain turbulence through the magnetorotational instability (MRI) \citep{balbus91}. 
It certainly appears to be the case that the instability could operate in dead zones \citep{gammie96}, and regions where non ideal MHD effects such as ambipolar diffusion act to moderate the strength of MRI turbulence \citep{bai15}. 
Precisely how the SWI operates under these conditions will require high resolution MHD simulations of spiral waves propagating in magnetized disks. 

In addition to understanding how the inclusion of additional physical processes could modify the SWI, future work is needed to examine how the SWI changes our understanding of astrophysical phenomena where spiral waves play an important role. 
The list of these includes the following: \\
-- Spiral wave propagation in disks with external binary companions, such as protoplanetary disks, CV systems and X-ray binaries. 
Issues of interest include how efficiently the spiral waves are damped by the SWI and how this affects the wave amplitude as a function of position in the disk, and whether or not the turbulent stresses associated with the SWI modify the outburst cycles associated with CV systems. 
In recent work, \citet{ju16} showed that spiral shocks can co-exist with MRI turbulence in CV systems, so it will be interesting to investigate applications of the SWI in the systems.\\
-- FU Orionis outbursts driven by spiral waves excited by self-gravity in disks around young stars.
The idea here is that spiral waves launched in the outer disk during the infall phase can heat the inner disk sufficiently to ionize it, and drive accretion onto the star by switching on the MRI \citep{zhu10, bae14}. 
We came across the SWI while conducting 3-D simulations designed to address this very issue, and in a forthcoming paper we will present a study that examines the viability of this picture.\\
-- Planet formation in disks with external binary companions. 
Numerous exoplanets are known to orbit one star that is a member of a binary system, perhaps the most famous of these being $\gamma$ Cephei \citep{hatzes03}. 
The excitation of spiral waves in a protoplanetary disk by an external binary is already known to have a strong influence on planet formation \citep{kley08, paardekooper08}. 
The excitation of turbulence by the SWI will also have an important influence, particularly on the settling of dust grains and the growth of small particles. \\
-- Formation of circumbinary planets. 
The situation regarding the growth of the SWI for spiral waves that propagate outwards from an outer Lindblad resonance is different than for an inward propagating wave. Ignoring the effects of buoyancy, and noting that inertial wave frequencies must lie in the range $0 < \omega_{\rm i} \le \Omega$, we infer that the resonant excitation of inertial waves by an inner companion occurs in a relatively narrow range of radii where the doppler-shifted frequency of the spiral $\omega_{\rm s} \simeq \omega_{\rm i}/2$. 
A fluid element at large radius in a disk sees a large value for $\omega_{\rm s}$, such that the resonance condition cannot be satisfied, unless buoyancy forces are also included in a disk where the Brunt-V\"as\"al\"a frequency is large. 
This suggests that there may be a range of radii in circumbinary disks where the SWI operates, making it difficult for small particles to grow there, and for planets to form in situ. The details will depend on the disk thermodynamics. \\
-- Formation of planets in the presence of a giant planet. It is generally believed that Jupiter was the first planet to form in the Solar System, and its early presence has been invoked to explain a number of physical and dynamical features that are observed today, such as the dynamical state of the asteroid belt and the small mass of Mars \citep{walsh11}. 
It will be of interest to examine how the onset of the SWI, induced by the spiral waves excited by a growing Jupiter, influenced the orbits of asteroids though stochastic forcing, and modified the growth of the terrestrial planets or their precursor embryos.
The influence could be strong if this occurred mainly through chondrule/pebble accretion, as explored recently by \citet{johansen15} and \citet{levison15}.

\section{CONCLUSION}
\label{sec:conclusion}
We have presented the results of high resolution, 3-D simulations of circumstellar disks that are perturbed by two-armed spiral density waves, and we have shown that a broad range of disk models are subject to a parametric instability involving the excitation of pairs of inertial waves that interact resonantly with the spiral wave. 
This spiral wave instability (SWI) gives rise to turbulence that transports angular momentum and causes vertical mixing. 
The apparent robustness of the SWI under changes to physical conditions in the disks suggests that it may arise in a broad range of astrophysical settings. 
Future work is required to understand and evaluate its influence on the physical evolution and observational appearance of these systems.

\acknowledgments

Authors thank the anonymous referee for a helpful report that improved the initial manuscript. 
J.B. thanks Fred Adams and Steve Lubow for valuable conversations, and Charles Gammie for insightful comments on the initial manuscript.
This research was supported in part through computational resources and services provided by Advanced Research Computing at the University of Michigan, Ann Arbor, and by HPC resources provided by IT Services at Queen Mary University of London.
This work used the Extreme Science and Engineering Discovery Environment (XSEDE), which is supported by National Science Foundation grant number ACI-1053575.
The authors acknowledge the San Diego Supercomputer Center at University of California, San Diego and the Texas Advanced Computing Center at The University of Texas at Austin for providing HPC resources that have contributed to the research results reported within this paper. 
This work used the DiRAC Complexity system, operated by the University of Leicester IT Services, which forms part of the STFC DiRAC HPC Facility (www.dirac.ac.uk). The equipment is funded by BIS National E-Infrastructure capital grant ST/K000373/1 and STFC Operations grant ST/K0003259/1. DiRAC is part of the national E-Infrastructure.

\appendix

\section{VERTICALLY STRATIFIED, LOCALLY ISOTHERMAL MODELS}

We introduce locally isothermal disk models, in which a radial temperature gradient is imposed by assuming $H_0/R_0=0.05$ and $q=-1$, and an isothermal equation of state is adopted.
With this setup, the disk aspect ratio maintains a constant value at all radii. Since the radial temperature gradient produces a non-zero vertical gradient in the disk rotation (see Equation \ref{eqn:init_vel}), this model is susceptible to the vertical shear instability \citep{nelson13}. We implement the density slope $p=-1.5$, so the initial disk structure is identical to the T1R-0 and T1R-0-3D models of \citet{nelson13}. The model parameters are summarized in Table \ref{tab:viso_parameters}.

\begin{figure}
\centering
\epsscale{1.2}
\plotone{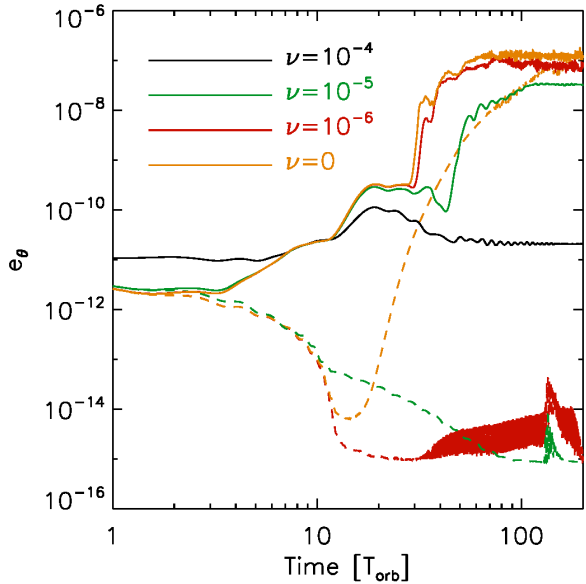}
\caption{Time evolution of the meridional kinetic energy $e_\theta$ for vertically isothermal models with different kinematic viscosities. The solid curves show the results with the spiral potential amplitude of $\mathcal{A}=5.0\times10^{-4}$, while the dashed curves show the results with $\mathcal{A}=0$ to check whether the vertical shear instability develops in the absence of spiral waves. The vertical shear instability is suppressed with $\nu \geq 10^{-6}$, which is in good agreement with \citet[][see their Figure 9]{nelson13}. The spiral wave instability is suppressed with $\nu=10^{-4}$ as in the globally isothermal models.}
\label{fig:kinz_viso}
\end{figure}

In order to check whether or not the vertical shear instability develops in the absence of imposed spiral waves, we first run calculations with the spiral potential turned off ($\Phi_p=0$).
We vary the kinematic viscosity using the values $\nu=0, 10^{-6}, 10^{-5}$.
As inferred from the time evolution of the integrated meridional kinetic energy presented in Figure \ref{fig:kinz_viso}, the vertical shear instability develops when $\nu=0$ but is suppressed with $\nu \geq 10^{-6}$, which is in good agreement with \cite{nelson13}.
The small fluctuations at $t \gtrsim 30$ in the VISO-V6 model and at $t \gtrsim 130$ in the VISO-V5 model are signatures of the vertical shear instability, but note that they do not grow over time and damp out because of viscous dissipation.

Having confirmed that the disk is stable to the vertical shear instability with $\nu \geq 10^{-6}$, we now introduce spiral waves into the disk.
We test with four different kinematic viscosity values: $\nu=0, 10^{-6}, 10^{-5}$, and $10^{-4}$.
As shown in Figure \ref{fig:kinz_viso}, the result is consistent with the globally isothermal models: the spiral wave instability is triggered with $\nu \leq 10^{-5}$ and is completely suppressed with $\nu = 10^{-4}$.
Also, the saturated energy level is smaller with larger viscosity, as expected.

We run the $\nu=10^{-6}$ model with the four codes introduced in Section \ref{sec:codes}.
As we discussed earlier, the detailed evolution is not identical because of the difference in dissipative properties of the codes, and the fact that NIRVANA and INABA3D use uniform radial meshes, whereas FARGO3D and PLUTO use logarithmically spaced grid cells. 
Nevertheless, the results show reasonably good agreement to each other, and all codes reproduce the SWI.

\begin{deluxetable*}{lccccccccccc}
\tablecolumns{15}
\tabletypesize{\tiny}
\tablecaption{Locally Isothermal Model Parameters \label{tab:viso_parameters}}
\tablewidth{0pt}
\tablehead{
\colhead{Run label} & 
\colhead{Code} & 
\colhead{Numerical Resolution} & 
\colhead{$q$} &
\colhead{$\mathcal{A}$} & 
\colhead{Kinematic} & 
\colhead{$\gamma$} & \\
\colhead{} & 
\colhead{} & 
\colhead{($N_r \times N_\phi \times N_\theta$ )} & 
\colhead{} & 
\colhead{} & 
\colhead{Viscosity $\nu$} &
\colhead{} 
 }
\startdata
VISO-V0 & FARGO3D & $512\times128\times128$ & -1 & $5.0\times10^{-4}$  & 0 &  1.0 \\
VISO-V6 & FARGO3D & $512\times128\times128$ & -1 & $5.0\times10^{-4}$  & $10^{-6}$ &  1.0 \\
VISO-V5 & FARGO3D & $512\times128\times128$ & -1 & $5.0\times10^{-4}$  & $10^{-5}$ &  1.0 \\
VISO-V4 & FARGO3D & $512\times128\times128$ & -1 & $5.0\times10^{-4}$  & $10^{-4}$ &  1.0 \\
\hline
VISO-N & NIRVANA & $3200\times128\times128$ & -1 & $5.0\times10^{-4}$  & $10^{-6}$ &  1.0 \\
VISO-P & PLUTO & $512\times128\times128$ & -1 & $5.0\times10^{-4}$  & $10^{-6}$ &  1.0 \\
VISO-I & INABA3D & $3200\times128\times128$ & -1 & $5.0\times10^{-4}$  & $10^{-6}$ &  1.0
\enddata
\end{deluxetable*}

\begin{figure}
\centering
\epsscale{1.2}
\plotone{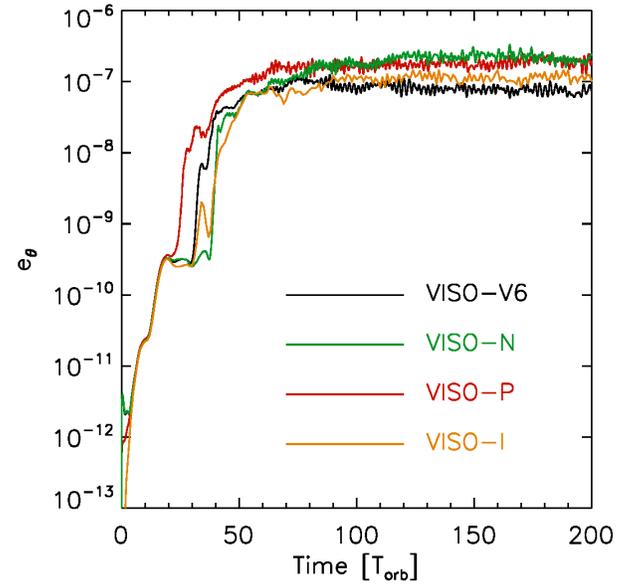}
\caption{Time evolution of the meridional kinetic energy $e_\theta$ for the vertically isothermal models with different codes (VISO-V6, VISO-N, VISO-P, VISO-I).}
\label{fig:viso-comp}
\end{figure}

\end{document}